\documentclass{sig-alternate-05-2015}

\newif\ifsubmit\submittrue

\usepackage{graphics}
\usepackage{url}
\usepackage[override]{cmtt}
\usepackage[pdftex,hyperref,svgnames]{xcolor}
\usepackage{rotating}
\usepackage{multirow}
\usepackage{pdfsync}
\usepackage{xcolor}
\usepackage{flushend}

\usepackage{xspace}
\usepackage{booktabs}
\usepackage{subcaption}

\usepackage[pdftex,plainpages=false]{hyperref}

\ifsubmit
\newcommand{\mwh}[1]{}
\newcommand{\awr}[1]{}
\newcommand{\jp}[1]{}
\newcommand{\pxm}[1]{}
\newcommand{\michelle}[1]{}
\newcommand{\dml}[1]{}
\else
\newcommand{\mwh}[1]{\textcolor{blue}{MWH -- #1}}
\newcommand{\awr}[1]{\textcolor{green}{AWR -- #1}}
\newcommand{\jp}[1]{\textcolor{BlueViolet}{JP -- #1}}
\newcommand{\pxm}[1]{\textcolor{orange}{PM -- #1}}
\newcommand{\michelle}[1]{\textcolor{teal}{MLM -- #1}}
\newcommand{\dml}[1]{\textcolor{purple}{dml -- #1}}
\fi

\let\oldfootnote\footnote
\renewcommand{\footnote}[1]{\oldfootnote{\small #1}}

\newcommand{\bibifi}[0]{BIBIFI\xspace}
\CopyrightYear{2016}
\setcopyright{acmlicensed}
\conferenceinfo{CCS'16,}{October 24 -- 28, 2016, Vienna, Austria}
\isbn{978-1-4503-4139-4/16/10}\acmPrice{\$15.00}
\doi{http://dx.doi.org/10.1145/2976749.2978382}

\title{Build It, Break It, Fix It: Contesting Secure Development}

\author{
 Andrew Ruef~~~~~
 Michael Hicks~~~~~
 James Parker \\
 Dave Levin~~~~~
 Michelle L. Mazurek~~~~~
 Piotr Mardziel$^\dagger$ \\
 ~ \\
        \affaddr{University of Maryland} $\qquad$ $~^\dagger$\affaddr{Carnegie
          Mellon University}
}

\hypersetup{%
  pdftitle={Build It, Break It, Fix It: Contesting Secure Development},
  pdfauthor={Blinded for Submission},
  pdfkeywords={},
  bookmarksnumbered,
  pdfstartview={FitH},
  colorlinks,
  citecolor=black,
  filecolor=black,
  linkcolor=black,
  urlcolor=black,
  breaklinks=true,
}

\begin{document}

\maketitle

\begin{abstract}
  Typical security contests focus on breaking or mitigating the impact
  of buggy systems. We present the Build-it, Break-it, Fix-it
  (\bibifi) contest, which aims to assess the ability to securely
  build software, not just break it. In \bibifi, teams build specified
  software with the goal of maximizing correctness, performance, and
  security. The latter is tested when teams attempt to break other
  teams' submissions. Winners are chosen from among the best builders
  and the best breakers. \bibifi was designed to be open-ended---teams
  can use any language, tool, process, etc.~that they like. As such,
  contest outcomes shed light on factors that correlate with
  successfully building secure software and breaking insecure
  software. During 2015, we ran three contests involving a total of 116
  teams and two different programming problems. Quantitative analysis
  from these contests found that the most efficient build-it
  submissions used C/C++, but submissions coded in other statically-typed
  languages were less likely to have a security flaw; build-it teams
  with diverse programming-language knowledge also produced more
  secure code. Shorter programs correlated with better
  scores. Break-it teams that were also successful build-it teams were
  significantly better at finding security bugs.
\end{abstract}

\bigskip
\noindent
% \mwh{Add keywords? Classification stuff is not required.}
% \mwh{We were using sig-alternate ``hacked'' -- why? I reverted it for now}

%intro the idea and concept of games for security
\section{Introduction}

Cybersecurity
contests~\cite{mdc3,nationalccdc,defconctf,csawctf,cansec}
are popular proving grounds for cybersecurity
talent. Existing contests largely focus on \emph{breaking} (e.g.,
exploiting vulnerabilities or misconfigurations) and
\emph{mitigation} (e.g., rapid patching or reconfiguration). They do
not, however, test contestants' ability to \emph{build} (i.e., design
and implement) systems that are secure in the first place. Typical
programming contests~\cite{topcoder,acmprogramming,icfpcontest} do
focus on design and implementation, but generally ignore
security. This state of affairs is unfortunate because experts have
long advocated that achieving security in a computer system requires
treating security as a first-order design goal~\cite{SaltzerSc75}, and
is not something that can be added after the fact. As such, we should not
assume that good breakers will necessarily be good
builders~\cite{marketplace-defcon}, nor that top coders necessarily produce
secure systems.

This paper presents \textbf{Build-it, Break-it, Fix-it} (\bibifi),
a new security contest with a focus on \emph{building secure
  systems}.
A \bibifi contest has three phases. The first phase, \emph{Build-it},
asks small development teams to build software according to a provided
specification that includes security goals. The software is scored for
being correct, efficient, and feature-ful. The second phase,
\emph{Break-it}, asks teams to find defects in other teams' build-it
submissions. Reported defects, proved via test cases vetted by an
oracle implementation, benefit a
break-it team's score and penalize the build-it team's score; more
points are assigned to security-relevant problems. (A team's break-it
and build-it scores are independent, with prizes for top scorers in
each category.)  The final phase, \emph{Fix-it}, asks builders to fix
bugs and thereby get points back if the process discovers that distinct
break-it test cases identify the same defect.

\bibifi's design aims to
minimize the manual effort of running a contest, helping it scale. \bibifi's
structure and scoring system also aim to
encourage meaningful outcomes, e.g., to ensure that the top-scoring
build-it teams really produce secure and efficient software. Behaviors
that would thwart such outcomes are discouraged. For example, break-it
teams may submit a limited number of bug reports per build-it
submission, and will lose points during fix-it for test cases that
expose the same underlying defect or a defect also identified by other
teams. As such, they are encouraged to look for bugs broadly (in many
submissions) and deeply (to uncover hard-to-find bugs).

In addition to providing a novel educational experience, \bibifi
presents an opportunity to study the building and breaking process
scientifically. In particular, \bibifi contests may serve as a
quasi-controlled experiment that correlates participation data with
final outcome. By examining artifacts and participant surveys, we can
study how the choice of build-it programming language, team size and
experience, code size, testing technique, etc.~can influence a team's
(non)success in the build-it or break-it phases. To the extent
that contest problems are realistic and contest participants represent
the professional developer community, the results of this study may
provide useful empirical evidence for practices that help or harm
real-world security. Indeed, the contest environment could be used to
incubate ideas to improve development security, with the best ideas
making their way to practice.
% \mwh{Bobby comment (incorporated above): Might be interesting to start
%   with some sort of an observation that a competitive environment may
%   serve as a good incubator? for developing security critical code.
%   This reflects how the code gets deployed in the real-world, and the
%   bibifi framework provides opportunity for iterative refinement of
%   software, again reflecting what happens in the wild.}

This paper studies the outcomes of three \bibifi contests that we held
during 2015, involving two different programming problems. The first
contest asked participants to build a \emph{secure, append-only log}
for adding and querying events generated by a hypothetical art gallery
security system. Attackers with direct access to the log, but lacking
an ``authentication token,'' should not be able to steal or corrupt
the data it contains. The second and third contests were run
simultaneously. They asked participants to build a pair of
\emph{secure, communicating programs}, one representing an ATM and the
other representing a bank. Attackers acting as a man in the middle (MITM)
should neither be able to steal information (e.g., bank account names
or balances) nor corrupt it (e.g., stealing from or adding money to
accounts). Two of the three contests drew participants from a MOOC
(Massive Online Open Courseware) course on cybersecurity. These
participants (278 total, comprising 109 teams) had an average of
10 years of programming experience and had just completed a
four-course sequence including courses on secure software and
cryptography. The third contest involved U.S.-based graduate and
undergraduate students (23 total, comprising 6 teams) with 
less experience and training.

\bibifi's design permitted it to scale reasonably well. For example,
one full-time person and two part-time judges ran the first 2015
contest in its entirety. This contest involved 156 participants
comprising 68 teams, which submitted more than 20,000 test cases. And
yet, organizer effort was limited to judging whether the few hundred
submitted fixes addressed only a single conceptual defect; other work
was handled automatically or by the participants themselves. 

Rigorous quantitative analysis of the contests' outcomes revealed several
interesting, statistically significant effects. Considering build-it
scores: Writing code in C/C++ increased build-it scores initially, but
also increased chances of a security bug found later. Interestingly, the increased
insecurity for C/C++ programs appears to be almost entirely attributable to 
memory-safety bugs. Teams that had
broader programming language knowledge or that wrote less code also
produced more secure implementations. Considering break-it scores:
Larger teams found more bugs during the break-it phase. Greater
programming experience and knowledge of C were also
helpful. Break-it teams that also qualified during the build-it
phase were significantly more likely to find a security bug than those
that did not. Use of advanced tools such as fuzzing or static analysis 
did not provide a significant advantage among our contest participants.

We manually examined both build-it and break-it
artifacts. Successful build-it teams typically employed third-party
libraries---e.g., SSL, NaCL, and BouncyCastle---to implement
cryptographic operations and/or communications, which freed up worry
of proper use of randomness, nonces, etc. Unsuccessful teams 
typically failed to employ cryptography, implemented it incorrectly,
used insufficient randomness, or failed to use
authentication. Break-it
teams found clever ways to exploit security problems;
some MITM implementations were quite sophisticated.

In summary, this paper makes two main contributions. First, it
presents \bibifi, a new security contest that encourages building, not
just breaking. Second, it presents a detailed description of three
\bibifi contests along with both a quantitative and qualitative
analysis of the results. We will be making the \bibifi code and
infrastructure publicly available so that others may run their own
competitions; we hope that this opens up a line of research built on
empirical experiments with secure programming
methodologies.\footnote{This paper subsumes a previously published
  short workshop paper~\cite{bibifi-cset15} and a short invited
  article~\cite{ruef15bibifiNW}. The initial \bibifi design and
  implementation also appeared in those papers, as did a brief
  description of a pilot run of the contest. This paper presents many
  more details about the contest setup along with a quantitative and
  qualitative analysis of the outcomes of several larger contests.}
More information, data, and opportunities to participate are available at
\url{https://builditbreakit.org}.

The rest of this paper is organized as follows.
We present the design of \bibifi in~\S\ref{sec:contest} and describe
specifics of the contests we ran in~\S\ref{sec:contests}.
We present the quantitative analysis of the data we collected from
these contests in~\S\ref{sec:analysis}, and qualitative analysis
in~\S\ref{sec:stories}.
We review related work in~\S\ref{sec:related} and conclude
in~\S\ref{sec:conclusions}.

\section{Build-it, Break-it, Fix-it} % {{{
\label{sec:contest}

This section describes the goals, design, and implementation of the
\bibifi competition.
At the highest level, our aim is to create an environment that closely
reflects real-world development goals and constraints, and to encourage
build-it teams to write the most secure code they can, and break-it
teams to perform the most thorough, creative analysis of others' code
they can.
We achieve this through a careful design of how the competition is run
and how various acts are scored (or penalized). We also aim to
minimize the manual work required of the organizers---to allow the
contest to scale---by employing automation and proper participant
incentives.

\subsection{Competition phases} % {{{
\label{sec:design}

We begin by describing the high-level mechanics of what occurs during a
\bibifi competition.
\bibifi may be administered on-line, rather than on-site, so teams may
be geographically distributed. The contest comprises three phases, each
of which last about two weeks for the contests we describe in this
paper.
% \mwh{Show picture of these phases?}

\bibifi begins with the \textbf{build-it phase}.  Registered
build-it teams aim to implement the target software system according to a
published specification created by the contest organizers. A
suitable target is one that can be completed by good programmers in a
short time (just about two weeks, for the contests we ran), is easily
benchmarked for performance, and has an interesting attack surface. The
software should have specific security goals---e.g., protecting private
information or communications---which could be compromised by poor
design and/or implementation. The software should also not be too
similar to existing software to ensure that contestants do the
coding themselves (while still taking advantage of high-quality libraries and
frameworks to the extent possible). The software must build and run on a standard Linux VM made
available prior to the start of the contest.  Teams must develop using
Git~\cite{git}; with each push, the contest infrastructure downloads
the submission, builds it, tests it (for correctness and performance),
and updates the scoreboard. \S\ref{sec:contests} describes the
two target problems we developed: (1)~an append-only log; and
(2)~a pair of communicating programs that simulate a bank and an ATM. 

% While the participants are implementing the target (in whatever
% programming language they like, as long as it is installed on the
% given VM image), we will record observations about their activities.
% In particular, we will store their software version history as
% development proceeds (e.g., in a repository we will
% host), and we will periodically inquire (using a browser-based
% popup~\cite{barrett,consolvo}) as to their current activities.
% When finished, contestants submit their code to the contest submission
% system, which will test performance and correctness inside an isolated
% virtual machine.
% If the submission passes all the tests, we accept it and then
% benchmark it for total performance, scoring it according to how much
% better it does than the minimum baseline, up to a maximum of 200
% points for the fastest submission.\footnote{These point values should
%   be taken with a grain of salt; we discuss how they should be set in
%  ~\S\ref{sec:contest-threats}.}
% Resubmissions are acceptable up until the deadline.

% The choice of programming task and the length of phase one are both
% important parameters of the contest, and we discuss the rationale for
% particular choices in detail in~\S\ref{sec:parameters}.  For the
% sake of discussion, assume phase one is 48 hours (so contestants
% complete the task in a (busy) weekend), and that the task is a parser
% for an obscure file format that dumps statistics into a database.

The next phase is the \textbf{break-it phase}.
% We will also ensure (e.g., by manual inspection and running
% Moss~\cite{moss}) that multiple teams have not submitted largely the
% same code.  Then the submissions are organized according to meta-data
% such as programming language, framework, build system, etc.
Break-it teams can download, build, and inspect all qualifying build-it
submissions, including source code; to qualify, the submission must
build properly, pass all correctness tests, and not be purposely
obfuscated (accusations of obfuscation are manually judged by the
contest organizers).  % \pxm{you mean only ``qualifying'' build-it submissions, see next comment about lack of ``qualifying'' definition} 
% \pxm{Why is obfuscation not a
%   valid strategy, as related to the freedom comment on the first page?
% Perhaps add a footnote and forward reference to why obfuscation is not
% allowed?} \mwh{I'm inclined to just do nothing.}
We randomize each break-it team's
view of the build-it teams' submissions,\footnote{This avoids spurious
unfair effects, such as if break-it teams investigating code in the
order in which we give it to them.} but organize them by meta-data,
such as programming language.
When they think they have found a defect, breakers submit a test case
that exposes the defect and an explanation of the issue.
To encourage coverage, a break-it team may only submit up a fixed
number of test cases per build-it submission.
\bibifi's infrastructure automatically judges whether a submitted test
case truly reveals a defect. For example, for a correctness bug, it
will run the test against a reference implementation (``the oracle'')
and the targeted submission, and only if the test passes on the former
but fails on the latter will it be accepted.\footnote{Teams
can also earn points by reporting bugs in the oracle, i.e., where its
behavior contradicts the written specification; these reports are
judged by the organizers.}
%   Any failing test is potentially worth points; we take the view that
%   a correctness bug is potentially a security bug waiting to be
%   exploited.  Many bugs that seem benign on the surface (e.g.,
%   integer overflows and race conditions) have ultimately led to
%   security compromises in surprising ways.
%
More points are awarded to bugs that clearly reveal security problems,
which may be demonstrated using alternative test formats. The
auto-judgment approaches we developed for the two different contest
problems are described in~\S\ref{sec:contests}. 

% which is defined as a failing test that causes the program to produce
% an output it should never produce.
% For example, a test case that causes the program to write to the
% console ``I'm pwned'' might be considered an exploit.
% Break-it participants are limited to submitting no more than 10
% failing test cases per build-it submission.
% (They are free to provide the same test case for multiple submissions,
% if they like.)

The final phase is the \textbf{fix-it phase}.
% During the week that precedes it, the organizers will analyze the
% submitted test cases, and attempt to judge whether any of them are
% faulty (i.e., the test case is wrong, not the software) or are
% duplicates (i.e., they are just variations that trigger the same
% underlying defect).
% We will develop automated techniques, described in
% ~\S\mref{sec:auto-grading}, to help with this task.
% At the start of the fix-it phase, the organizers 
Build-it teams are provided with the bug reports and test cases
implicating their submission. They may fix flaws these test cases
identify; if a single fix corrects more than one failing test case, the
test cases are ``morally the same,'' and thus points are only deducted
for one of them.  The organizers determine, based on information
provided by the build-it teams and other assessment, whether a
submitted fix is ``atomic'' in the sense that it corrects only one
conceptual flaw; if not, the fix is rejected.
% At the end of this phase, we will have identified a likely set of
% unique, correct test cases.

%Each of these phases can vary in duration from a weekend to a couple of weeks,
%depending on the contest's target audience (and how much time they
%have available). 
% Inspired by the ICFP programming contest~\cite{icfpcontest}, we had
% originally set the contest to run on three consecutive (3-day)
% weekends, but found that this was too aggressive (too few teams
% qualified)~\cite{bibifi-cset15}. 

Once the final phase concludes, prizes are awarded to the best
builders and best breakers as determined by the scoring
system described next.

% }}}

\subsection{Competition scoring} % {{{
\label{sec:contest-goals}

\bibifi's scoring system aims to encourage the contest's basic goals,
which are that the winners of the build-it phase truly
produced the highest quality software, and that the winners of the
break-it phase performed the most thorough, creative analysis of
others' code. The scoring rules create incentives for good behavior
(and disincentives for bad behavior). 

\subsubsection{Build-it scores} % {{{
To reflect real-world development concerns, the winning build-it team
would ideally develop software that is correct, secure, and efficient.
While security is of primary interest to our contest,
developers in practice must balance these other aspects of quality
against security~\cite{cfi,ulfar-personal}, leading to a set of
trade-offs that cannot be ignored if we wish to understand real
developer decision-making.

To encourage these, each build-it team's score is the sum of the
\emph{ship} score\footnote{The name is meant to evoke a quality
  measure at the time software is shipped.} and the \emph{resilience} score.
The ship score is composed of points gained for correctness tests and
performance tests.  Each mandatory correctness test is worth $M$
points, for some constant $M$, while each optional correctness test is
worth $M/2$ points.
Each performance test has a numeric measure depending on the specific
nature of the programming project---e.g., latency, space consumed,
files left unprocessed---where lower measures are better.
A test's worth is $M \cdot{} (\mathit{worst} - v) / (\mathit{worst} -
\mathit{best})$, where $v$ is the measured result, $\mathit{best}$ is
the measure for the best-performing submission, and $\mathit{worst}$ is
the worst performing. As such, each performance test's value ranges
from 0 to $M$. 

The resilience score is determined after the break-it and fix-it
phases, at which point the set of unique defects against a submission
is known. For each \emph{unique} bug found against a team's submission,
we subtract $P$ points from its resilience score; as such, it is
non-positive, and the best possible resilience score is $0$. For
correctness bugs, we set $P$ to $M/2$; for crashes that violate memory
safety we set $P$ to $M$, and for exploits and other security property
failures we set $P$ to $2M$. 
% \pxm{Is ``memory safety'' a universally understood term? Are all
%   crashes examples of memory safety issue?} \mwh{Yes, and no (propose
%   we do nothing)}

% They earn more points the more correctness tests they pass and the
% better performance they gain.
%
% Moreover, it is in the build-it teams' best interests to write
% maintainable (i.e., understandable) code because they can gain points
% back by fixing flaws. \pxm{Fixing only allows them to try to prove how
%   many different bugs are actually the same bug. Are you suggesting
%   maintainable code has fewer ``unique'' bugs relative to not
%   maintainable code?}
% }}}

\subsubsection{Break-it scores} % {{{
Our primary goal with break-it teams is to encourage them to find as
many defects as possible in the submitted software, as this would give
greater confidence in our assessment that one build-it team's
software is of higher quality than another's.  While we are
particularly interested in obvious security defects, correctness
defects are also important, as they can have non-obvious security
implications. % \pxm{Also, writing software without some necessary
  % features means the software is smaller and hence less likely to have
  % bugs of all kinds.} \mwh{So? (Propose we do nothing)}

After the break-it phase, a break-it team's score is the summed value
of all defects they have found, using the above $P$ valuations.  After
the fix-it phase, this score is reduced.  In particular, each of the
$N$ break-it teams' scores that identified the same defect are adjusted
to receive $P/N$ points for that defect, splitting the $P$ points among
them.  % Dividing this adjusted
% break-it score by the unadjusted score yields a \emph{uniqueness
% coefficient}, where values closer to 1 indicate a propensity to find
% bugs undetected by other teams.

Through a combination of requiring concrete test cases and scoring,
\bibifi encourages break-it teams to follow the spirit of the
competition.
First, by requiring them to provide test cases as evidence of a defect
or vulnerability, we ensure they are providing useful bug reports.  By
providing $4\times$ more points for security-relevant bugs than for
correctness bugs, we nudge break-it teams to look for these sorts of
flaws, and to not just focus on correctness issues. (But a different
ratio might work better; see \S\ref{sec:limitations}.) Because break-it
teams are limited to a fixed number of test cases per submission, and
because they could lose points during the fix-it phase for submitting
test cases that could be considered ``morally the same,'' break-it
teams are encouraged to submit tests that demonstrate different bugs. 
% discouraged from submitting many tests they suspect are ``morally
% the same''; as they could lose points for them during the fix-it phase
% they are better off submitting tests demonstrating different bugs.
Limiting per-submission test cases also encourages examining many
submissions. Finally, because points for defects found by other teams
are shared, break-it teams are encouraged to look for hard-to-find
bugs, rather than just low-hanging fruit.
%  Together, these incentives encourage both broad and deep exploration
%  to find many unique bugs.

\subsection{Discussion}

The contest's design also aims to enable scalability by reducing work
on contest organizers. In our experience, \bibifi's design succeeds at
what it sets out to achieve, but is not perfect. We close by
discussing some limitations.

\subsubsection{Minimizing manual effort}
Once the contest begins, manual effort by the organizers is, by design,
limited. All bug reports submitted during the break-it phase are
automatically judged by the oracle; organizers only need to vet any
bug reports against the oracle itself. Organizers may also need to judge
accusations by breakers of code obfuscation by builders. Finally,
organizers must judge whether submitted fixes address a single defect;
this is the most time-consuming task. It is necessary because we
cannot automatically determine whether multiple bug reports against
one team map to the same software defect. Instead, we incentivize
build-it teams to demonstrate overlap through fixes; organizers
manually confirm that each fix addresses only a single defect, not
several.

Previewing some of the results presented later, we can confirm that
the design works reasonably well. For example, as detailed in
Table~\ref{tab:bugs-fixes}, 68 teams submitted 24,796 test cases in
our Spring 2015 contest. The oracle auto-rejected 15,314 of these, and
build-it teams addressed 2,252 of those remaining with 375 fixes, a
6$\times$ reduction. Most confirmations that a fix truly addressed a
single bug took 1--2 minutes each. Only 30 of these fixes were
rejected. No accusations of code obfuscation were made by break-it
teams, and few bug reports were submitted against the oracle. All
told, the Spring 2015 contest was successfully managed by one
full-time person, with two others helping with judging.

\subsubsection{Limitations}
\label{sec:limitations}

While we believe \bibifi's structural and scoring incentives are
properly designed, we should emphasize several limitations. 

First, there is no guarantee that all implementation defects will be
found. Break-it teams may lack the time or skill to find problems in
all submissions, and not all submissions may receive equal
scrutiny. Break-it teams may also act contrary to incentives and focus
on easy-to-find and/or duplicated bugs, rather than the harder and/or
unique ones. % \pxm{How is the previous a limitation if the scoring is
  % designed specifically to prevent it???} \mwh{But people don't always
  % act rationally; I bring this up because in fact that happened in our
  % contest runs.}
Finally, break-it teams may find defects that the \bibifi
infrastructure cannot automatically validate, meaning those defects
will go unreported. However, with a large enough pool of break-it
teams, and sufficiently general defect validations automation, we still
anticipate good coverage both in breadth and depth.

Second, builders may fail to fix bugs in a manner that is in their
best interests. For example, in not wanting to have a fix rejected as
addressing more than one conceptual defect, teams may use several
specific fixes when a more general fix would have been allowed. 
Additionally, teams that are out of contention for prizes may
simply not participate in the fix-it phase.\footnote{Hiding scores
  during the contest might help mitigate this, but would harm incentives during break-it
  to go after submissions with no bugs reported against them.}
 We observed this behavior
for our contests, as described in~\S\ref{sec:resilience}. Both
actions decrease a team's resilience score (and correspondingly
increase breakers' scores). We can mitigate these issues with
sufficiently strong incentives, e.g., by offering prizes to all
participants commensurate with their final score, rather than offering
prizes only to the top scorers.

Finally, there are several design points in the problem definition
that may skew results. For example, too few correctness tests may
leave too many correctness bugs to be found during break-it
(distracting break-it teams' attention from security issues); too many
correctness tests may leave too few (meaning teams are differentiated
insufficiently by general bug-finding ability). Scoring prioritizes
security problems 4 to 1 over correctness problems, but it is hard to
say what ratio makes the most sense when trying to maximize real-world
outcomes; both higher and lower ratios could be argued. Finally, performance
tests may fail to expose important design trade-offs (e.g., space
vs.~time), affecting the ways that teams approach maximizing their 
ship scores. For the contests we report in this paper, we are
fairly comfortable with these design points. In particular, our
earlier contest~\cite{bibifi-cset15} prioritized security bugs 2-to-1
and had fewer interesting performance tests, and outcomes were better
when we increased the ratio.

% }}}

\subsubsection{Discouraging collusion} % {{{
\bibifi contestants may form teams however they wish, and may
participate remotely.
This encourages wider participation, but it also opens the possibility
of collusion between teams, as there cannot be a judge overseeing their
communication and coordination.
There are three broad possibilities for collusion, each of which 
\bibifi's scoring discourages.

First, two break-it teams could consider sharing bugs they find with
one another.  By scaling the points each finder of a particular bug
obtains, we remove incentive for them to both submit the same bugs, as
they would risk diluting how many points they both obtain.
%One team could take a ``king-maker'' approach and simply give all bugs
%to the other team and not submit any themselves, but this is in
%essence equivalent to having formed a single team, and thus does not
%violate the rules.

The second class of collusion is between a build-it team and a
break-it team, but neither have incentive to assist one another.
The zero-sum nature of the scoring between breakers and builders
places them at odds with one another; revealing a bug to a break-it
team hurts the builder, and not reporting a bug hurts the breaker.

Finally, two build-it teams could collude, for instance by sharing code
with one another. It might be in their interests to do this in the
event that the competition offers prizes to two or more build-it teams,
since collusion could obtain more than one prize-position.  We use
judging and automated tools (and feedback from break-it teams) to
detect if two teams share the same code (and disqualify them), but it
is not clear how to detect whether two teams provided out-of-band
feedback to one another prior to submitting code (e.g., by holding
their own informal ``break-it'' and ``fix-it'' stages).  We view this
as a minor threat to validity; at the surface, such assistance appears
unfair, but it is not clear that it is contrary to the goals of the
contest, that is, to develop secure code.

% \mwh{Should we mention the collusion scenario that Stu Schechter
%   raised, and why it's unlikely? Thinking no ...}

% }}}

% }}}

\subsection{Implementation} % {{{
\label{sec:implementation}

\begin{figure}[t!]
  \begin{centering}
    %\subimport{figures/}{summary_system}
	\includegraphics[width=\columnwidth]{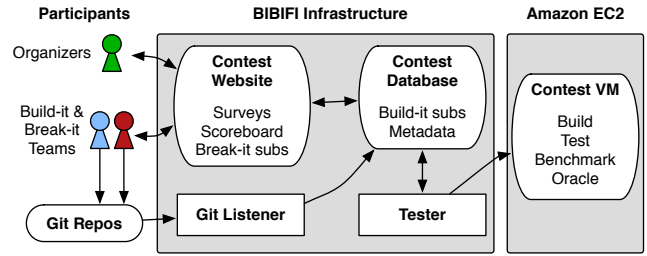}
  \end{centering}
  \caption{\label{fig:system} Overview of \bibifi's implementation.
\vspace{-2ex}}
\end{figure}

%% Figure~\ref{fig:system} illustrates components of the contest
%% implementation (along the left), and the interactions between
%% participants and outside elements (on the right). 

% \bibifi is designed to support many participating teams without
% requiring significant manual effort by the organizers. 

% to several hundred participating teams
% within a given contest.  The more teams we have, the greater our impact
% from both a teaching and research point of view: on the teaching side,
% more students experience the desired learning outcomes, and on the
% research side, we have a greater corpus of data to draw on.
% %
% To this end, we have built an infrastructure that reduces the amount of
% work for the organizers through best-effort automation.
% %
% We describe the details our infrastructure's design and implementation
% here; 
Figure~\ref{fig:system} provides an overview of the \bibifi
implementation. It consists of a web frontend, providing the interface
to both participants and organizers, and a backend for testing builds
and breaks. Two key goals of the infrastructure are security---we do not
want participants to succeed by hacking \bibifi itself---and
scalability. 

\paragraph*{Web frontend} % {{{

%% The locus of the \bibifi contest is the web application.
% Double blind: running at \url{https://builditbreakit.org}.
%
Contestants sign up for the contest through our web application
frontend, and fill out a survey when doing so, to gather demographic
and other data potentially relevant to the contest outcome (e.g.,
programming experience and security training). During the contest, the
web application tests build-it submissions and break-it bug reports,
keeps the current scores updated, and provides a workbench for the
judges for considering whether or not a submitted fix covers one bug or
not.

To secure the web application against unscrupulous participants, we
implemented it in ${\sim}$11,000 lines of Haskell using the Yesod
\cite{yesodweb} web framework backed by a PostgreSQL \cite{psql}
database. Haskell's strong type system defends against use-after-free,
buffer overrun, and other memory safety-based attacks.  The use of
Yesod adds further automatic protection against various attacks like
CSRF, XSS, and SQL injection. As one further layer of defense, the web
application incorporates the information flow control framework
LMonad~\cite{parker2014lmonad}, which is derived from LIO~\cite{lio},
in order to protect against inadvertent information leaks and privilege
escalations. LMonad dynamically guarantees that users can only access
their own information. 
%In addition, it enforces policies that only administrators can perform
%privileged actions like creating new announcements.

% to view scores and indicate where their code is stored. The web 
% application orchestrates the compilation of submitted code and the
% execution of test cases for all phases of the contest. The web 
% application also allows the organizers to make announcements and
% post links to static assets such as the specification and virtual
% machine images.
% In addition, the web application provides an interface where 
% judges can report judgments of submissions when required. 

% }}}

\paragraph*{Testing backend} % {{{

The backend infrastructure is used during the build-it phase to test
for correctness and performance, and during the break-it phase to
assess potential vulnerabilities. It consists of ${\sim}$5,100 lines of
Haskell code (and a little Python). 
%and about 900 more shared with the
%web application.

To automate testing, we require contestants to specify a URL to a
Git~\cite{git} repository hosted on either Github or Bitbucket, and
shared with a designated \texttt{bibifi} username, read-only. The
backend ``listens'' to each contestant repository for pushes, upon
which it downloads and archives each
commit. Testing is then handled by a scheduler that spins up an Amazon
EC2 virtual machine which builds and tests each submission. We require
that teams' code builds and runs, without any network access, in an
Ubuntu Linux VM that we share in advance. Teams can request that we
install additional packages not present on the VM. The use of VMs
supports both scalability (Amazon EC2 instances are dynamically
provisioned) and security (using fresh VM instances prevents a team
from affecting the results of future tests, or of tests on other teams'
submissions). 

All qualifying build-it submissions may be downloaded by break-it
teams at the start of the break-it phase.  % \pxm{how is qualification tested? problem descriptions below are ambiguous regarding this as they state features must be implemented, but not how this is checked.}
%
%% % Already mentioned above
%% The web application randomizes the order the submissions are presented
%% to each break-it team to encourage broad coverage. 
%
As break-it teams identify bugs, they prepare a (JSON-based) file
specifying the buggy submission along with a sequence of commands with
expected outputs that demonstrate the bug. Break-it teams commit and push this
file (to their Git repository). The backend uses the file
to set up a test of the implicated submission to see if it indeed is a
bug.

\section{Contest Problems} % {{{
\label{sec:contests}

This section presents the two programming problems we developed for
the contests held during 2015, including prob\-lem-specific notions of
security defect and how breaks exploiting such defects are
automatically judged.
%
% While the competitions differ in substance, they both have potential
% for bugs in correctness, privacy, and integrity.
% %
% This overlap permits a more direct comparison of the competitions in
% our analysis in~\S\ref{sec:analysis}.

\newpage
\subsection{Secure log (Spring 2015)}
\label{sec:problem}

The secure log problem was motivated as support for an art gallery
security system. Contestants write two programs. The first,
\texttt{logappend}, appends events to the log; these events indicate
when employees and visitors enter and exit gallery rooms. The second,
\texttt{logread}, queries the log about past events. To qualify,
submissions must implement two basic queries (involving the current
state of the gallery and the movements of particular individuals), but they
could implement two more for extra points (involving time spent in the
museum, and intersections among different individuals' histories). An
empty log is created by \texttt{logappend} with a given authentication
token, and later calls to \texttt{logappend} and
\texttt{logread} on the same log must use that token or the
requests will be denied.

A canonical way of implementing the secure log is to treat the
authentication token as a symmetric key for authenticated
encryption, e.g., using a combination of AES and HMAC. % \pxm{Could use reference
  % or explanation of the prior sentence.} 
  %       \jp{These are standard crypto algorithms. I'm not sure they need references.}
%% to both encrypt the log, e.g., using
%% AES, and to sign it, e.g., using HMAC with SHA-256.
%
There are several tempting shortcuts that we anticipated build-it
teams would take (and that break-it teams would exploit).
For instance, one may be tempted to encrypt and sign individual log records
as opposed to the entire log, thereby making \texttt{logappend} faster.
But this could permit integrity breaks that duplicate or reorder log
records.
Teams may also be tempted to implement their own
encryption rather than use existing libraries, or to simply sidestep
encryption altogether.
\S\ref{sec:stories} reports several cases we observed.

%% One temptation is to encrypt/MAC individual log records, to make
%% \texttt{logappend} faster. But this could permit an adversary to
%% duplicate or re-order records in the log, allowing possible integrity
%% attacks. A wise approach would be to use an existing encryption
%% library, e.g., SSL or NaCl, rather than rolling one's own.
%% %
%% \mwh{Cite other pitfalls, e.g., simply using encryption, but not
%% signing.}

A submission's performance is measured in terms of time to perform a
particular sequence of operations, and space consumed by the resulting
log. Correctness (and \emph{crash}) bug reports comprise sequences of
\texttt{logread} and/or \texttt{logappend} operations with expected
outputs (vetted by the oracle).
Security is defined by \emph{privacy} and \emph{integrity}: any attempt to learn something
about the log's contents, or to change them, without the use of the
\texttt{logread} and \texttt{logappend} and the proper token should be
disallowed. How violations of these properties are specified and tested
is described next.

% \paragraph*{Correctness and crash bugs}
% Break-it teams demonstrate correctness bugs in a submission $S$
% by providing a test in a JSON-based format which defines a transcript of
% \texttt{logread} and/or \texttt{logappend} operations and their
% expected outputs. If the oracle (our canonical solution) produces the
% expected outputs when executing these operations but submission $S$ does not, a
% correctness violation is confirmed. If submission $S$ crashes with a
% memory error (e.g., segfault, bus error, etc.) while executing the
% transcript then a security-relevant \emph{crash} failure is confirmed.

% \medskip
% Privacy and integrity bugs can occur without appearing to be a
% correctness or crash bug; automating validation of such therefore
% requires extra consideration.

\paragraph*{Privacy breaks}
%
%
%% For privacy or integrity bugs, we might consider accepting
%% text-based explanations of why a bug could violate security. Or we
%% could require an exploit, e.g., as a script that runs a
%% series of commands (\texttt{logread}, \texttt{logappend}, and others
%% to directly manipulate logs, say) and shows the extraction or
%% corruption of information. Both approaches have scalability problems
%% as they rely on human judgment (e.g., to 
%% be convinced the script really is showing a problem). 
%
%% To avoid these
%% problems, we took the following approach.
%
When providing a build-it submission to the break-it teams, we also
included a set of log files that were generated using a sequence of
invocations of that submission's \texttt{logappend} program. We
generated different logs for different build-it submissions, using a
distinct command sequence and authentication token for each. All logs
were distributed to break-it teams without the authentication token;
some were distributed without revealing the sequence of commands (the
``transcript'') that generated them. For these, a break-it team could
submit a test case involving a call to \texttt{logread} (with the
authentication token omitted) that queries the file. The \bibifi
infrastructure would run the query on the specified file with the
authentication token, and if the output matched that specified by the
breaker, then a privacy violation is confirmed.

\paragraph*{Integrity breaks}
For about half of the generated log files we also provided the
transcript of the \texttt{logappend} operations (\emph{sans} auth
token) used to generate the
file. A team could submit a test case specifying the name of the log file, the
contents of a corrupted version of that file, and a \texttt{logread}
query over it (without the authentication token). For both the
specified log file and the corrupted one, the \bibifi infrastructure
would run the query using the correct authentication token. An
integrity violation is detected if the query command produces a
non-error answer for the corrupted log that 
differs from the correct answer (which can be confirmed against the
transcript using the oracle).

%% Why? A correct implementation would return an error when given a
%% corrupted log file. If the breaker submitted the original log file as
%% ``corrupted,'' it would not produce an error, but its answer would
%% obviously match that of the original. If one of the other legal
%% log files was submitted, it should also be rejected as it was generated
%% with a different auth token than the original. 

This approach to determining privacy and integrity breaks has the
%benefit and 
drawback that it does not reveal the \emph{source} of the issue, only
that there is (at least) one. As such, we cannot automatically tell two
privacy or two integrity breaks apart.
We sidestep this issue by counting only up to one integrity break and
one privacy break against the score of each build-it submission, even
if there are multiple defects that could be exploited to produce
privacy/integrity violations.

% \mwh{Past/present tense in the following is perhaps not what it should be here. Dave
  % and/or Michelle could check ...}

\subsection{Securing ATM interactions (Fall 2015)}
\label{sec:atm-problem}

The ATM problem asks builders to construct two communicating programs:
\texttt{atm} acts as an ATM client, allowing customers to set up an
account, and deposit and withdraw money, while \texttt{bank} is a server that processes client
requests, tracking bank balances. \texttt{atm} and \texttt{bank} should
only permit a customer with a correct \emph{card file} to learn or
modify the balance of their account, and only in an appropriate way
(e.g., they may not withdraw more money than they have). In addition,
\texttt{atm} and \texttt{bank} should only communicate if they can
authenticate each other. They can use an \emph{auth file} for this
purpose; it will be shared between the two via a
trusted channel unavailable to the attacker.\footnote{In a real
deployment, this might be done by ``burning'' the auth file into the ATM's
ROM prior to installing it.}  Since the \texttt{atm} is communicating
with \texttt{bank} over the network, a ``man in the middle'' (MITM)
could observe and modify exchanged messages, or insert new messages.
The MITM could try to compromise security despite not having access to
auth or card files.

% \pxm{I don't think this belongs in problem summary:} \mwh{Why? We do
%   it for the previous problem and you didn't complain. I feel it
%   grounds the reader.}
A canonical way of implementing the
\texttt{atm} and \texttt{bank} 
programs would be to use public key-based authenticated and
encrypted communications. The auth file could be used as the \texttt{bank}'s public
key to ensure that key negotiation initiated by the \texttt{atm} is
with the \texttt{bank} and not the MITM. When creating an account, the
card file should be a suitably large random number, so that the MITM is
unable to feasibly predict it. It is also necessary to protect against
replay attacks by using nonces or similar mechanisms.
As with the secure log, a wise approach
would be use a library like OpenSSL to implement these features. Both
good and bad implementations we observed in the competition are
discussed further in~\S\ref{sec:stories}. 
% \pxm{Perhaps move this whole paragraph to~\S\ref{sec:stories}.}

Build-it submissions' performance is measured as the time to complete
a series of benchmarks involving various \texttt{atm}/\texttt{bank} interactions.\footnote{This
transcript was always serial, so there was no direct motivation to
support parallelism for higher throughput.} Correctness 
(and \emph{crash}) bug reports comprise sequences of
\texttt{atm} commands where the targeted submission produces
different outputs than the oracle (or crashes).
Security defects are specified as follows.
%% As with the log example,
%% security boils down to both privacy and integrity. The former is
%% violated if a MITM attacker can discover the name of an account holder,
%% the amount in a transaction, or an account's balance.  The latter is
%% violated if the MITM can modify an account holder's balance.

% \paragraph*{Correctness bugs} 

% To prove a correctness problem in submission $S$, a breaker could
% submit a transcript of commands to the \texttt{atm} program. These
% commands are first run with submission $S$'s \texttt{atm} and
% \texttt{bank}, then with the oracle's corresponding pair of programs. If either of $S$'s \texttt{atm} or \texttt{bank}
% crashed with a memory error during the test, that would constitute a
% security-relevant \emph{crash} failure. 

\paragraph*{Integrity breaks} 

Integrity violations are demonstrated using a custom MITM program
that acts as a proxy: It listens on a specified IP address and TCP
port,\footnote{All submissions were required to communicate via TCP.}
and accepts a connection from the \texttt{atm} while connecting to the
\texttt{bank}. The MITM program can thus observe and/or modify
communications between \texttt{atm} and \texttt{bank}, as well as drop
messages or initiate its own. We provided a Python-based proxy as a
starter MITM: It sets up the connections and forwards
communications between the two endpoints.

To demonstrate an integrity violation, the MITM sends requests to a
\emph{command server}. It can tell the
server to run inputs on the \texttt{atm} and it can ask for the
card file for any account whose creation it initiated. Eventually the MITM
will declare the test complete. At this point, the same set of
\texttt{atm} commands is run using the oracle's \texttt{atm} and
\texttt{bank} \emph{without the MITM}. This means that any messages
that the MITM sends directly to the target submission's \texttt{atm} or \texttt{bank}
will not be replayed/sent to the oracle.
If the oracle and target both
complete the command list without error, but they differ on the
outputs of one or more commands, or on the balances of
accounts at the bank whose card files were not revealed to the MITM
during the test, then there is evidence of an integrity violation.

As an example (based on a real attack we observed), consider a
submission that uses deterministic encryption without nonces in
messages. The MITM could direct the command server to withdraw money
from an account, and then replay the message it observes. When run on
the vulnerable submission, this would debit the account twice. But
when run on the oracle without the MITM, no message is replayed,
leading to differing final account balances. A correct submission
would reject the replayed message, which would invalidate the break.

\paragraph*{Privacy breaks} 

Privacy violations are also demonstrated using a MITM. In this case, at
the start of a test the command server will generate two random
values, ``amount'' and ``account name.'' If by the end of the test no
errors have occurred and the attacker can prove it knows the actual
value of either secret (by sending a command that specifies it),
the break is considered successful. Before
demonstrating knowledge of the secret, the MITM can
send commands to the server with a \emph{symbolic} ``amount''
and ``account name''; the server fills in the
actual secrets before forwarding these messages. The command server does
not automatically create a secret account or an account with a secret
balance; it is up to the breaker to do that (referencing the secrets
symbolically when doing so). 

As an example, suppose the target does not encrypt exchanged messages. Then a
privacy attack might be for the MITM to direct the command server to
create an account whose balance contains the secret amount. Then the
MITM can observe an unencrypted message sent from \texttt{atm} to
\texttt{bank}; this message will contain the actual amount, filled in
by the command server. The MITM can then send its guess to the command
server showing that it knows the amount. 

As with the log problem, we cannot tell whether an integrity or
privacy test is exploiting the same underlying weakness in a
submission, so we only accept one violation of each category against
each submission.

\paragraph*{Timeouts and denial of service}

One difficulty with our use of a MITM is that we cannot reliably detect
bugs in submissions that would result in infinite loops, missed
messages, or corrupted messages. This is because such bugs can be
simulated by the MITM by dropping or corrupting messages it
receives. Since the builders are free to implement any protocol they
like, our auto-testing infrastructure cannot tell if a protocol
error or timeout is due to a bug in the target or due to misbehavior of the MITM. As
such, we conservatively disallow reporting any such errors. Such flaws
in builder implementations might exist but evidence of those bugs might
not be realizable in our testing system. 
%\pxm{Miss some what? Is the thing you missed
%  something you'd want to discover but cannot?} \mwh{We should discuss}

% \subsection{Discussion}

% \mwh{work in?}

% Contest problems are necessarily limited in scope, and contest
% participants include students and non-professional developers; as
% such, results of our analysis may not translate to the real world. On
% the other hand: Many of our contestants have received significant
% training (the coursera sequence), and the contest problems reflect
% real-world problems (secure communication and storage) in a
% microcosm. Moreover, software is developed by a wide variety of
% contexts (think: cyber-physical systems/IOT), and our impression is
% that developers in these contexts may have similar levels of
% experience to those who participated in our contest.

% }}}

% \mwh{Add subsection on possible future problems here. E.g., targeting
%   the web (per reviewer 2), access control/authorization problems,
%   etc.?}

\section{Quantitative Analysis}
\label{sec:analysis}

This section analyzes data we have gathered from three contests we ran
during 2015.\footnote{We also ran a contest during Fall 2014~\cite{bibifi-cset15} but
  exclude it from consideration due to differences in how it was
  administered.}
We consider participants' performance in each phase of the
contest, including which factors contribute to high scores after the 
build-it round,
resistance to breaking by other teams, and strong performance as
breakers. 

We find that on average, teams that program in languages 
other than C and C++, and those whose members know more programming 
languages (perhaps a proxy for overall programming skill), are less 
likely to have security bugs identified in their code. However, when
memory management 
bugs are not included, programming language is no longer a significant factor, 
suggesting that memory safety is the main discriminator between C/C++ and 
other languages. Success in breaking, 
and particularly in identifying security bugs in other teams' code, is 
correlated with having more team members, as well as with
participating successfully in 
the build-it phase (and therefore having given thought to how to 
secure an implementation). Somewhat surprisingly, use of advanced techniques 
like fuzzing and static analysis did not appear to affect breaking success.  
Overall, integrity bugs were far more
common than privacy bugs or crashes. The Fall 2015 contest, which used 
the ATM problem, was associated with more security bugs than the 
Spring 2015 secure log contest. 

\subsection{Data collection}

For each team, we collected a variety of observed and self-reported
data. When signing up for the contest, teams reported standard
demographics and features such as coding experience and programming
language familiarity. After the contest, each team member optionally
completed a survey about their performance. In addition, we extracted
information about lines of code written, number of commits, etc.~from
teams' Git repositories.

Participant data was anonymized and stored in a manner approved by our
institution's human-subjects review board. Participants consented to
have data related to their activities collected, anonymized, stored,
and analyzed. 
A few participants did not consent to research involvement, 
so their personal data was not used in the data analysis. 
% \mwh{Say what we did with non-approved data}

\subsection{Analysis approach}
To examine factors that correlated with success in building and
breaking, we apply regression analysis. 
Each regression model attempts to explain some outcome variable using
one or more measured factors. 
For most outcomes, such as participants' scores, we can use ordinary
linear regression, which estimates how many points a given factor
contributes to (or takes away from) a team's score. 
To analyze binary outcomes, such as whether or not a security bug was
found, we apply logistic regression. 
This allows us to estimate how each factor impacts the likelihood of
an outcome. % \pxm{Why wouldn't a linear regression also give you an
  % estimate of impact of each factor?} \mwh{I think the statement is
  % clear; if you were curious you could look it up.}

We consider many variables that could potentially impact teams' results.
To avoid over-fitting, we initially select as potential factors those
variables that we believe are of most interest, within acceptable
limits for power and effect size. 
(Our choices are detailed below.) 
In addition, we test models with all possible combinations of these
initial factors and select the model with the minimum Akaike
Information Criterion (AIC)~\cite{burnham2011aic}. 
Only the final models are presented.

%We describe the results of each model below.
This was not a completely controlled experiment (e.g., we do not use
random assignment), so our models demonstrate correlation rather than
causation. Our observed effects may involve confounding variables, and many factors
used as independent variables in our data are correlated with each
other.  This analysis also assumes that the factors we examine have
linear effect on participants' scores (or on likelihood of binary
outcomes); while this may not be the case in reality, it is a common
simplification for considering the effects of many factors. We also
note that some of the data we analyze is self-reported, and thus may not be
entirely precise (e.g., some participants may have exaggerated about which
programming languages they know); however, minor deviations,
distributed across the population, act like noise and have little
impact on the regression outcomes. 

\subsection{Contestants}

We consider three contests offered at two times:

\textbf{Spring 2015}: We held one contest during May--June 2015
  as the capstone to a Cybersecurity MOOC sequence.\footnote{\url{https://www.coursera.org/specializations/cyber-security}} 
  Before competing in the capstone, participants passed courses on
  software security, cryptography, usable security, and hardware
  security. The contest problem was the secure log problem
  (\S\ref{sec:problem}).

\textbf{Fall 2015}: During Oct.--Nov.~2015, we offered two contests
  simultaneously, one as a MOOC capstone, and the other open to
  U.S.-based graduate and undergraduate
   students. We merged the contests after the build-it phase,
  due to low participation in the open contest; from here on we refer
  to these two as a single contest. The contest problem
  was the ATM problem (\S\ref{sec:atm-problem}).

The U.S.~had more contestants than any other country.  There was
representation from developed countries with a reputation both for high
technology and hacking acumen. Details of the most popular countries of
origin can be found in Table~\ref{tab:demo-country}, and additional
information about contestant demographics is presented in
Table~\ref{tab:demog}. 

% Participants in each contest used a variety of programming languages to 
% build their submissions. By examining their submissions, we determined the 
% primary language used by each team; Figure~\ref{fig:submissions-by-language} 
% illustrates the popularity of each language in each contest. % \pxm{where the two fall contests merged in this figure?}

\begin{table}[t]
\centering
\small
\begin{tabular}{lrrrrr}
\toprule
{\bf Contest}     & \textbf{USA} & \textbf{Brazil} & \textbf{Russia} &
\textbf{India} & \textbf{Other} \\
\midrule
Spring 2015 & 30  & 12    & 12     & 7      & 120   \\
Fall 2015   & 64  & 20    & 12     & 14     & 110  \\
\bottomrule
\end{tabular}
\caption{Contestants, by self-reported country.
\vspace{-2ex}}
\label{tab:demo-country}
\end{table}

\begin{table}[t]
\centering
\small
\label{my-label}
\begin{tabular}{@{}p{.35\columnwidth}|@{~}rrr@{}}
\toprule
\textbf{Contest} & \textbf{Spring '15}$^\dagger$ & \textbf{Fall '15}$^\dagger$ & \textbf{Fall '15} \\ 
\midrule
\small
{\small \# Contestants}                                       & 156            & 122                & 23                     \\
{\small \% Male}                                            & 91\% & 89\% & 100\%                \\
{\small \% Female}					&  5\% & 9\% & 0\% \\
{\small Age}                       & 34.8/20/61     & 33.5/19/69         & 25.1/17/31             \\
{\small \% with CS degrees}                                   & 35\%             & 38\%                 & 23\%                     \\
{\small Years programming}                                 & 9.6/0/30       & 9.9/0/37           & 6.6/2/13               \\
{\small \# Build-it teams}               & 61             & 34                 & 6                      \\
{\small Build-it team size}                                & 2.2/1/5        & 3.1/1/5            & 3.1/1/6                \\
{\small \# Break-it teams}               & 65           & 39                & 4                     \\
{\small (that also built)}              & (58)           & (32)                & (3)                     \\
{\small Break-it team size}                                & 2.4/1/5        & 3.0/1/5            & 3.5/1/6                \\
{\small \# PLs known per team}                & 6.8/1/22       & 10.0/2/20          & 4.2/1/8               \\
\bottomrule
\end{tabular}
\caption{Demographics of contestants from qualifying
  teams. $\dagger$ indicates MOOC participants. Some participants 
  declined to specify gender. Slashed values represent mean/min/max.}
\label{tab:demog}
\end{table}

%\begin{tabular}{lllllll}
%            & Number of Participants & M/F       & Mean Age & CS Degrees & Years Programming \\
%Spring 2015 & 191                    & 91\%/ 6\% & 34.7     & 50         & 5                 \\
%Fall 2015   & 230                    & 90\%/7\%  & 35.6     & 86         & 6                
%\end{tabular}

%\mwh{Put subsection here on the effectiveness of the automation? Take
%  paragraph from rebuttal. Can't think of a better place to put it.}

\subsection{Ship scores}
\label{ss:ship-it}

We first consider factors correlating with a team's \emph{ship} score,
which assesses their submission's quality
before it is attacked by the other teams (\S\ref{sec:design}).  This data set contains all
101 teams from the Spring 2015 and Fall 2015 contests that qualified
after the build-it phase. Both contests have nearly the same number of
correctness and performance tests, but different numbers of
participants. We set the constant multiplier $M$ to be 50
for both contests, which effectively normalizes the scores.

\paragraph*{Model setup}
  
To ensure enough power to find meaningful relationships, we decided to
aim for a prospective effect size roughly equivalent to Cohen's {\it
  medium} effect heuristic, $f^2=0.15$~\cite{Cohen:1988}. An effect 
  this size suggests the model can explain up to 13\% of the variance in 
  the outcome variable. 
With an assumed power of 0.75 and population $N=101$, we limited ourselves to nine
degrees of freedom, which yields a prospective $f^2=0.154$. (Observed
effect size for the final model is reported with the regression
results below.)  Within this limit, we selected the following potential
factors:
 
%Starting model:
%ship_score ~ problem + members_count + knowC + languageCount + 
%    coding_experience + loc + languageCat + coursera

\textbf{Contest:} Whether the team's submission was for the Spring 2015 contest
or the Fall 2015 contest.

\textbf{\# Team members:} A team's size.

\textbf{Knowledge of C:} The fraction of team members who listed C or C++ as a
programming language they know. We included this variable as a proxy
for comfort with low-level implementation details, a skill often
viewed as a prerequisite for successful secure building or breaking.

\textbf{\# Languages known:} How many unique programming languages
team members collectively claim to know (see the last row of
Table~\ref{tab:demog}). For example, on a two-member 
team where member A claims to know C++, Java, and Perl and member B
claims to know Java, Perl, Python, and Ruby, the language count would
be 5. 

\textbf{Coding experience:} The average years of programming
experience reported by a team's members.

\begin{figure}[t!]
\centering
  \includegraphics[width=0.95\columnwidth]{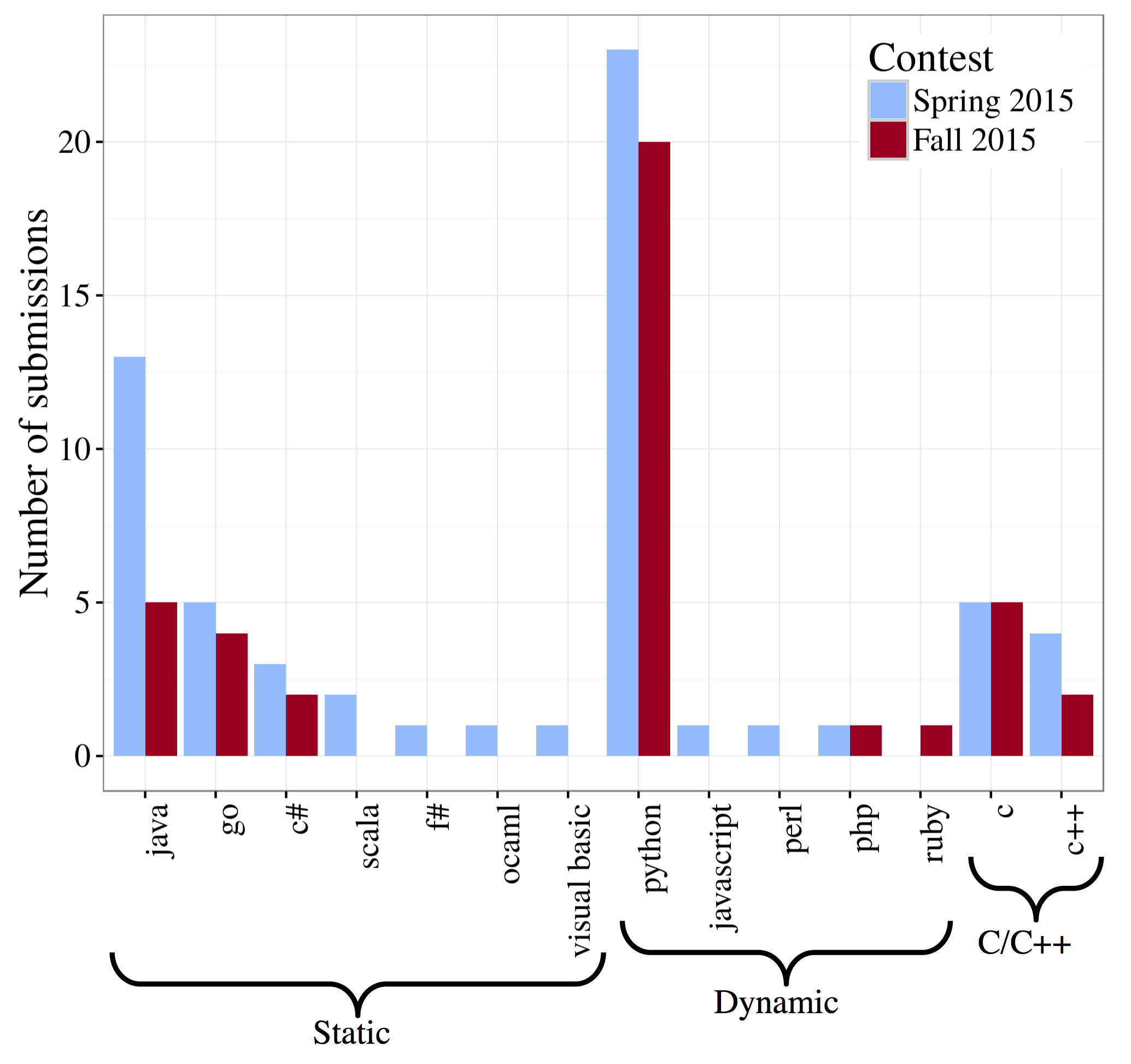}
\caption[]{The number of build-it submissions in each contest, organized by primary programming 
language used. The brackets group the languages into categories.}
\label{fig:submissions-by-language}
\end{figure}

\textbf{Language category:} We manually identified each team's submission
as having one ``primary" language. These languages were then assigned
to one of three categories: C/C++, statically-typed (e.g., Java, Go,
but not C/C++) and dy\-nam\-ically-typed (e.g., Perl, Python). C/C++ is the baseline
category for the regression.  Precise category allocations, and total
submissions for each language, segregated by contest, are given in
Figure~\ref{fig:submissions-by-language}.

\textbf{Lines of code:} The SLOC\footnote{\url{http://www.dwheeler.com/sloccount}} count
of lines of code for the team's final submission at qualification time.

\textbf{MOOC:} Whether the team was participating in the MOOC
capstone project. % \pxm{there were only 6 teams with ``false'', why select this?}

\paragraph*{Results}

\begin{figure}[t!]
\begin{center}
  \includegraphics[width=0.95\columnwidth]{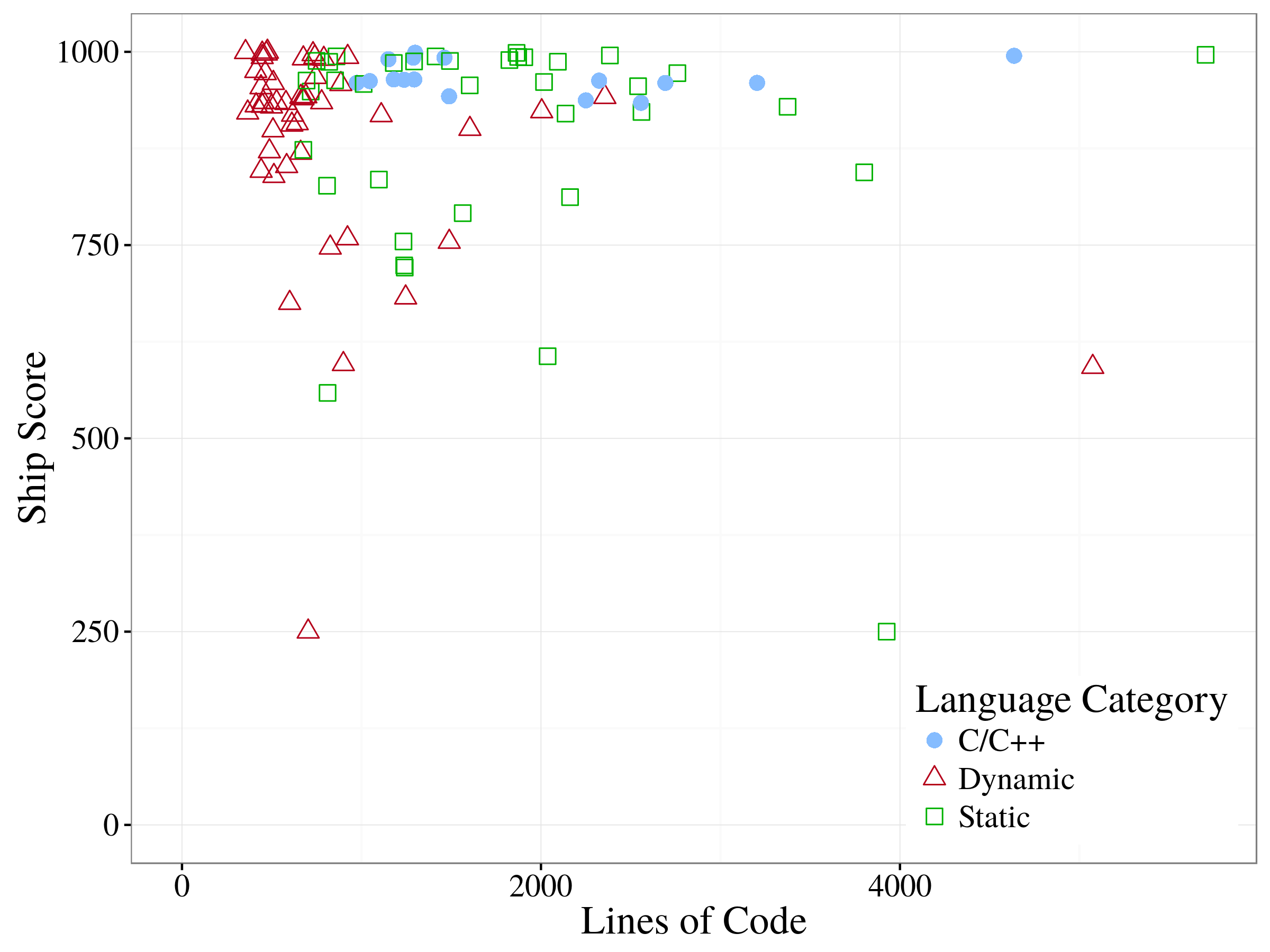}
\end{center}
\caption[]{Each team's ship score, compared to the lines of code in its implementation and 
organized by language category. Fewer LOC
and using C/C++ correlate with a higher ship score. }
\label{fig:LOC-langCat-ship}
\end{figure}

Our regression results (Table~\ref{tab:ship-model}) indicate that
ship score is strongly correlated with
language choice. Teams that programmed in C or C++ performed on
average 121 and 92 points better than those who programmed in
dynamically typed or statically typed languages,
respectively. Figure~\ref{fig:LOC-langCat-ship} illustrates that while
teams in many language categories performed well in this phase, only
teams that did not use C or C++ scored poorly. 

The high scores for C/C++ teams could be due to better scores on
performance tests and/or due to implementing optional features. We
confirmed the main cause is the former. 
Every C/C++ team for Spring 2015 implemented all optional
features, while six teams in the other categories implemented only 6
of 10, and one team implemented none; the Fall 2015 contest offered no
optional features. We artificially increased the scores of those seven
teams as if they had implemented all optional features and reran the
regression model. The resulting model had very similar coefficients. 

Our results also suggest that teams that were associated with the
MOOC capstone performed 119 points better than non-MOOC teams.
MOOC participants typically had more programming experience and CS
training.

Finally, we found that each additional line of code in a team's
submission was associated with a drop of 0.03 points in ship score. 
Based on our qualitative observations (see~\S\ref{sec:stories}), we
hypothesize this may relate to more reuse of code from libraries,
which frequently are not counted in a team's LOC (most libraries were installed directly
on the VM, not in the submission itself).  We also found that, as further noted
below, submissions that used libraries with more sophisticated,
lower-level interfaces tended to have more code and more
mistakes; their use required more code in the application, lending
themselves to missing steps or incorrect use, and thus security and
correctness bugs.
% i.e., more steps took place in the application (more code) but
% some steps were missed or carried out incorrectly (less secure/correct).
%
% \awr{I looked. There are two explanations. One is that higher level interfaces need less code, and teams using lower level interfaces did worse, because they would forget to authenticate or something. Another is that the more verbose code also just feels lower quality - there's a lot going on that could be tightened up. I feel pretty confident saying that judicious use of crypto libraries would both help your security score and depress your LOC count - you want the security code in your application to be one call.}
As shown in Figure~\ref{fig:LOC-langCat-ship}, LOC is also (as
expected) associated with the category of language being used. While
LOC varied widely within each language type, dynamic submissions were
generally shortest, followed by static submissions, and then those
written in C/C++ (which has the largest minimum size).

\begin{table}[t!]
%\begin{center}
\centering
\small
\begin{tabular}{l r r r }
\toprule
\textbf{Factor} & \textbf{Coef.} & \textbf{SE} & \textbf{$p$-value} \\
\midrule
\color{Gray} Fall 2015 & \color{Gray} -21.462 & \color{Gray} 28.359  & \color{Gray} 0.451~ \\
Lines of code & -0.031 & 0.014 & 0.036* \\
Dynamically typed & -120.577 & 40.953 & 0.004* \\
Statically typed & -91.782 & 39.388 & 0.022* \\
MOOC & 119.359 & 58.375 & 0.044* \\
\bottomrule
\end{tabular}
%\end{center}
%\vspace{-2ex}
\caption{Final linear regression model of teams' ship scores, indicating how many points 
each selected factor adds to the total score. % (The full set of 
% initial factors is listed in~\S\ref{ss:ship-it}; this model reflects minimum 
% AIC). Significant values are indicated with an asterisk.
Overall effect size  $f^2 =  0.163$.
\vspace{-3ex}} % \jp{Multiple R-squared:  0.1401,
                              % Adjusted R-squared:  0.09487} f2 = r2
                              % / 1 - r2 
\label{tab:ship-model}
%\vspace{-2ex}
\end{table}

%Graph and content to-do:
%\begin{itemize}
%\item verify that C did well due to performance (and not from extra features)
%\item boxplot of ship score, by language category (3 side-by-side boxplots)
%\item scatterplot: x=LOC, y=ship score, dots are colored/shaped to distinguish language category (e.g. blue circles for C, red triangles for static, etc.)
%\end{itemize}

\subsection{Code quality measures}
\label{sec:resilience}

Now we turn to measures of a build-it submission's quality---in terms
of its correctness and security---based on how it held up under
scrutiny by break-it teams.

\paragraph*{Resilience}
The total build-it score is the sum of ship score, just
discussed, and \emph{resilience}.  Resilience is a non-positive score
that derives from break-it teams' test cases that prove the presence of
defects. Builders may increase this score during the fix-it phase, as
fixes prevent double-counting test cases that identify the same defect
(see~\S\ref{sec:design}). % Therefore, we are interested in understanding
% the factors that correlate with resilience score.

\begin{figure}[t!]
\begin{centering}
\includegraphics[width=\columnwidth]{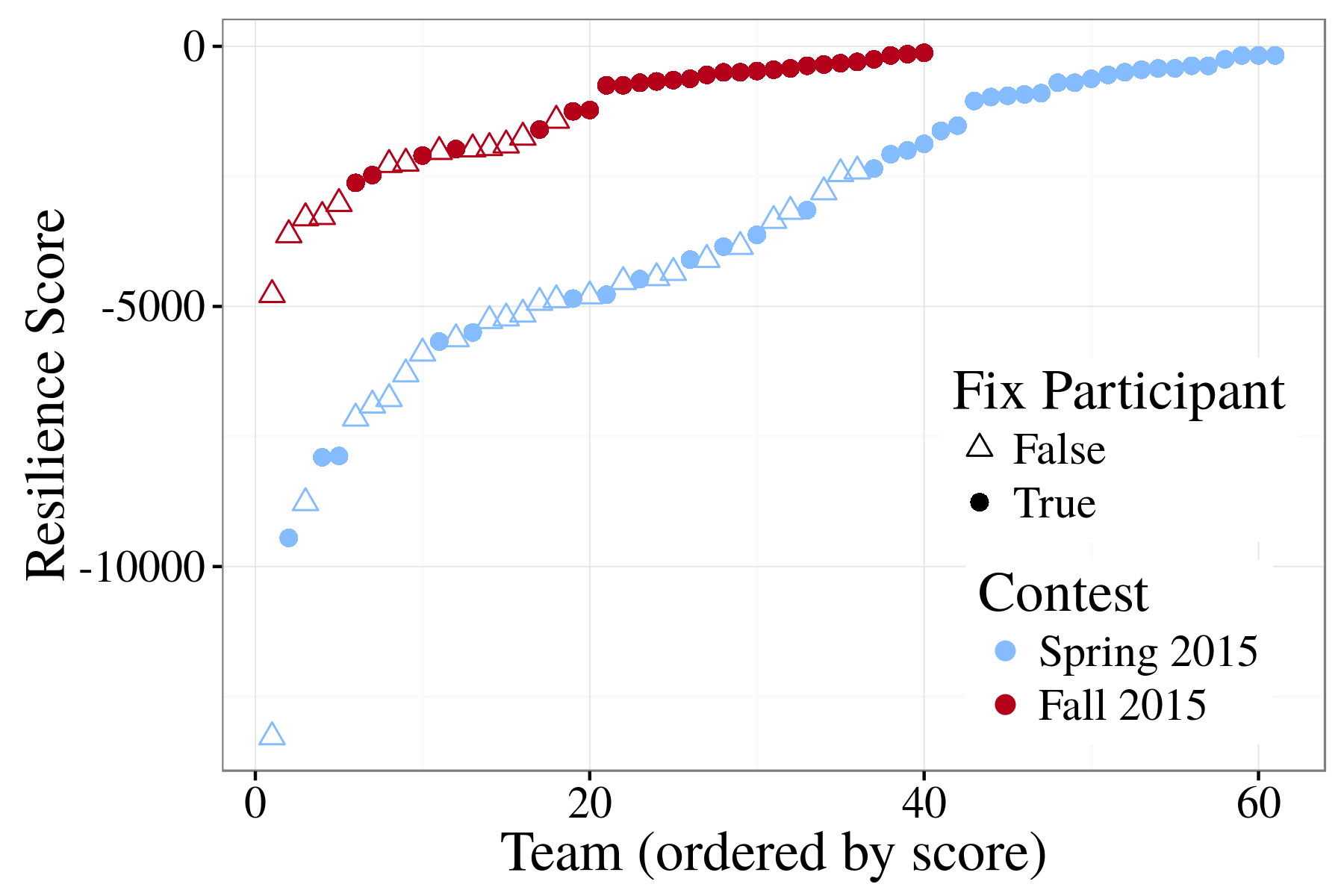}
\caption{Final resilience scores, ordered by team, and plotted for
  each contest problem. Build-it teams who did not bother to fix bugs
  generally had lower scores.}
\label{fig:p3}
\end{centering}
\end{figure}

Unfortunately, upon studying the data we found that a large percentage
of build-it teams opted not to fix any bugs reported against their code,
forgoing the scoring advantage of doing so. We can see this in
Figure~\ref{fig:p3}, which graphs the resilience scores (Y-axis) of
all teams, ordered by score, for the two contests. The circles in the
plot indicate teams that fixed at least one bug, whereas the triangles 
indicate teams that fixed no bugs. We can see that, overwhelmingly,
the teams with the lower resilience scores did not fix any bugs. We
further confirmed that fixing, or not, was a dominant factor by
running a regression on resilience score that included fix-it phase
participation as a factor (not shown). Overall, teams fixed an average 
of 34.5\% of bugs in Spring 2015 and 45.3\% of bugs in Fall 2015. 
Counting only ``active" fixers who fixed at least one bug, these averages 
were 56.9\% and 72.5\% respectively. 

\begin{table}[t]
\begin{center}
\small
\begin{tabular}{l r r }
\toprule
& \textbf{Spring 2015} & \textbf{Fall 2015} \\
\midrule
Bug reports submitted & 24,796 & 3,701 \\
Bug reports accepted & 9,482 & 2,482 \\
Fixes submitted & 375 & 166 \\
Bugs addressed by fixes & 2,252 & 966 \\
\bottomrule
\end{tabular}
\end{center}
\vspace{-2ex}
\caption{Break-it teams in each contest submitted bug reports, which were judged by the 
automated oracle. Build-it teams then submitted fixes, each of which could potentially 
address multiple bug reports.
 % (The full set of initial factors is listed in~\S\ref{ss:ship-it}; this model reflects minimum AIC). Significant values are 
% indicated with an asterisk.
%\vspace{-1ex}
}
\label{tab:bugs-fixes}
\vspace{-2ex}
\end{table}

Table~\ref{tab:bugs-fixes} digs a little further into the
situation. It shows that of the bug reports deemed acceptable by the
oracle (the second row), submitted fixes (row 3) addressed only 23\%
of those from  Spring 2015 and 38\% of those from Fall 2015 (row 4
divided by row 2). 

% We confirmed that not fixing bugs was a major contributor to overall
% score. We performed a linear regression of the resilience score using
% the same teams and factors described in~\S\ref{ss:ship-it},
% adding a boolean factor, \textbf{fix participant}, indicating whether
% the team fixed any bugs. The result is given in
% Table~\ref{tab:resilience-alt-model}. We can see that, overwhelmingly,
% the source of a team's final resilience score is whether it was a fix
% participant. % \pxm{Is it necessary to do a regression and add a table
%   % to show something I can infer using my eyeball on Figure 4?} 
% The percent of team members who know C appears in the 
% final model but is not significant, probably because it has such large 
% standard error. This suggests a large variance in score among teams 
% with similar degrees of C knowledge. 

This situation is disappointing, as we cannot treat resilience score as
a good measure of code quality (when added to ship score). Our
hypothesis is that participants were not sufficiently incentivized to
fix bugs, for two reasons. First, teams that are sufficiently far from
the lead may have chosen to fix no bugs because winning was
unlikely. Second, for MOOC students, once a minimum score is
achieved they were assured to pass; it may be that fixing (many) bugs was
unnecessary for attaining this minimum score. We are exploring
alternative reward structures that more strongly incentivize all teams to fix
all (duplicated) bugs. 

%round2_resilience_score ~ problem + members_count + knowC + languageCount + 
%    coding_experience + loc + languageCat + coursera + fix participant

% \begin{table}[t]
% \begin{center}
% \small
% \begin{tabular}{l r r r r}
% \toprule
% \textbf{Factor} & \textbf{Coef.} & \textbf{SE} & \textbf{$p$-value} \\
% \midrule
% Fall 2015 & 1918.0 & 393.7 & 0.001* \\
% \color{Gray} Knowledge of C & \color{Gray} 1078.0 & \color{Gray} 560.4 & \color{Gray} 0.057~ \\
% Fix participant & 2435.4 & 399.9 & <0.001* \\
% \bottomrule
% \end{tabular}
% \end{center}
% \vspace{-2ex}
% \caption{Final linear regression model of teams' resilience scores, indicating how many points 
% each selected factor adds to the total score. % (The full set of initial factors includes those listed in~\S\ref{ss:ship-it} plus a boolean factor indicating teams that fixed bugs; this model reflects 
% % minimum AIC). Significant values are indicated with an asterisk.
% Overall effect size $f^2 = 0.730$.
% % \jp{resilience-alt model. Multiple R-squared:  0.4219,  Adjusted R-squared:  0.404. $f^2 = 0.730$}
% }
% \label{tab:resilience-alt-model}
% \vspace{-2ex}
% \end{table}

\paragraph*{Presence of security bugs}

% \pxm{In \ref{sec:problem} and \ref{sec:atm-problem}, it sounded
%   like crashes were categorized as correctness bugs.}
While resilience score is not sufficiently meaningful, a useful alternative is
the likelihood that a build-it submission contains a security-relevant
bug; by this we mean any submission against which at least one 
crash, privacy, or integrity defect is demonstrated. In this model we used logistic regression
over the same set of factors as the ship model. 

Table~\ref{tab:secBug-model} lists the results of this logistic regression; 
the coefficients represent the change in log likelihood associated with each 
factor. Negative coefficients indicate lower likelihood of finding a security bug. 
For categorical factors, the exponential of the coefficient (Exp(coef)) indicates 
roughly how strongly that factor being true affects the likelihood relative to the baseline 
category.\footnote{In cases (such as the Fall 2015 contest) where the rate of 
security bug discovery is close to 100\%, the change in log likelihood starts to approach 
infinity, somewhat distorting this coefficient upwards.} For numeric factors, the 
exponential indicates how the likelihood changes with each unit change in that factor. 

\begin{table}[t]
\begin{center}
\small
\begin{tabular}{@{}l r r r r@{}}
\toprule
\textbf{Factor} & \textbf{Coef.} & \textbf{Exp(coef)} & \textbf{SE} & \textbf{$p$-value} \\
\midrule
Fall 2015 & 5.692 & 296.395 & 1.374 & <0.001* \\
\# Languages known & -0.184 & 0.832 & 0.086 & 0.033* \\
Lines of code & 0.001 & 1.001 & 0.0003 & 0.030* \\
\color{Gray} Dynamically typed & \color{Gray} -0.751 & \color{Gray} 0.472 & \color{Gray} 0.879 & \color{Gray} 0.393~ \\
Statically typed & -2.138 & 0.118 & 0.889 & 0.016* \\
\color{Gray} MOOC & \color{Gray} 2.872 & \color{Gray} 17.674 & \color{Gray} 1.672 & \color{Gray} 0.086~ \\
\bottomrule
\end{tabular}
\end{center}
\vspace{-2ex}
\caption{Final logistic regression model, measuring log likelihood of a security bug 
being found in a team's submission.
% (The full set of initial factors is listed in~\S\ref{ss:ship-it}; this model reflects minimum AIC). Significant values are 
% indicated with an asterisk.
\vspace{-6ex}
}
\label{tab:secBug-model}
\vspace{-2ex}
\end{table}

Fall 2015 implementations were $296\times$ as likely as Spring 2015
implementations to have a discovered security bug.\footnote{This 
coefficient is somewhat exaggerated (see prior footnote), but the 
difference between contests is large and significant.}  
We hypothesize this is due to the increased security design space in the
ATM problem as compared to the gallery problem. Although it is easier
to demonstrate a security error in the gallery problem, the ATM
problem allows for a much more powerful adversary (the MITM) that can
interact with the implementation; breakers often took advantage of
this capability, as discussed in~\S\ref{sec:stories}.

\begin{figure}[t]
\begin{centering}
\includegraphics[width=\columnwidth]{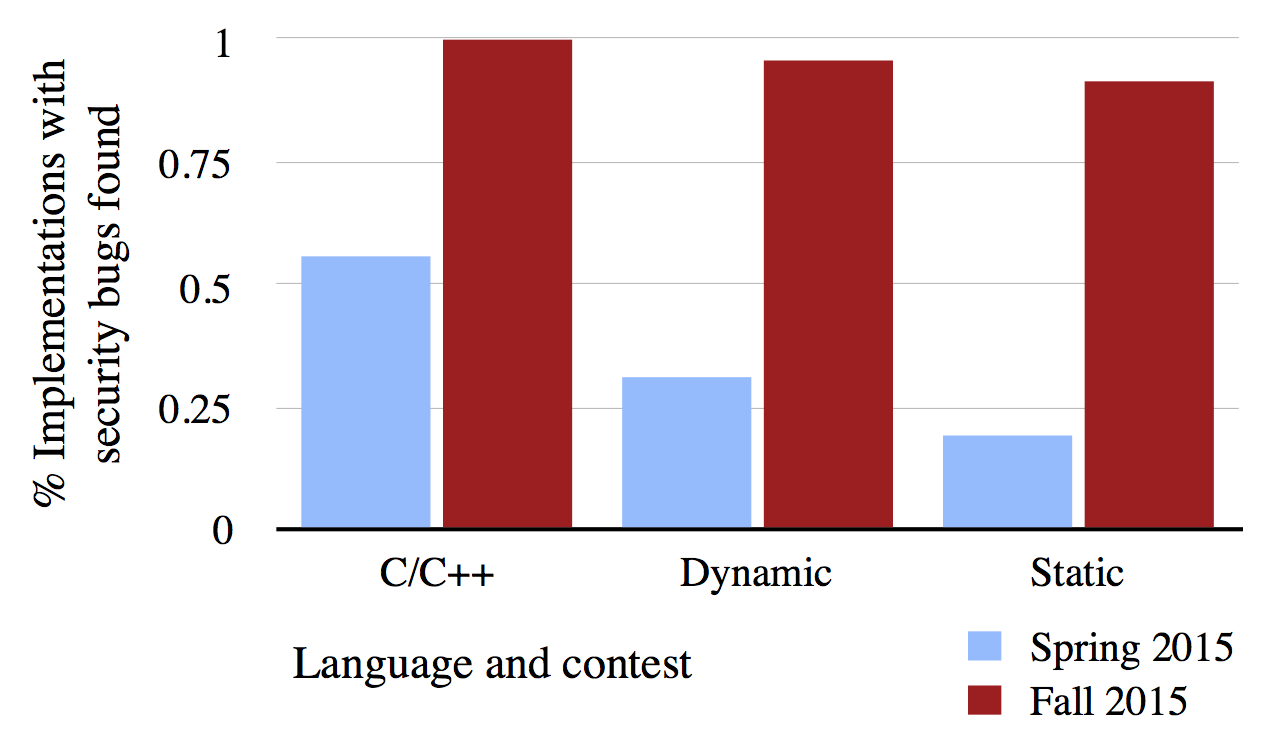}
\caption[]{The fraction of teams in whose submission a security bug was found, for 
each contest and language category.}
\label{fig:bugsfound-by-language}
\end{centering}
\end{figure}

The model also shows that C/C++ implementations were more likely to
contain an identified security bug than either static or dynamic
implementations. For static languages, this effect is significant and
indicates that a C/C++ program was about $8.5\times$ (that is,
$1/0.118$) as likely to contain an identified bug. This effect is
clear in Figure~\ref{fig:bugsfound-by-language}, which plots the
fraction of implementations that contain a security bug, broken down
by language type and contest problem. Of the 16 C/C++ submissions (see
Figure~\ref{fig:submissions-by-language}), 12 of them had a security bug: 5/9 for Spring 2015
and 7/7 for Fall 2015. All 5 of the buggy implementations from Spring
2015 had a crash defect, and this was the only
security-related problem for three of them; none of the Fall 2015 implementations had
crash defects. 

\begin{figure}[t]
\begin{centering}
\includegraphics[width=\columnwidth]{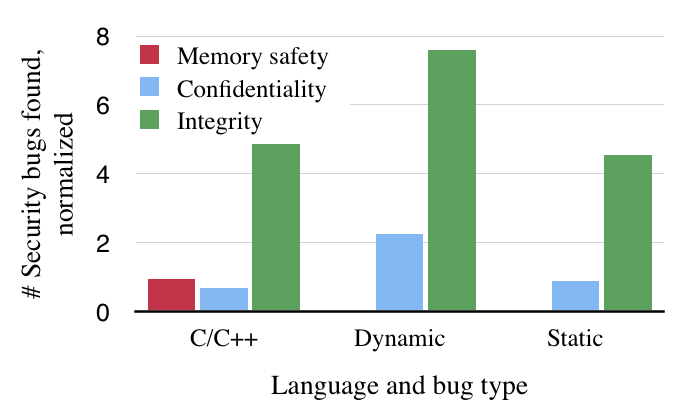}
\caption[]{How many of each type of security bug were found, across both contests, for each language category. 
Counts are normalized by the number of qualified Build-it submissions in each language category.}
\label{fig:bugcat-by-language}
\end{centering}
\end{figure}

If we reclassify crash defects as not security relevant
and rerun the model we find that the impact due to language category
is no longer statistically significant. This may indicate that lack of
memory safety is the main disadvantage to using C/C++ from a security
perspective, and thus a memory-safe C/C++ could be of significant
value. Figure~\ref{fig:bugcat-by-language} shows how many security bugs 
of each type (memory safety, integrity, privacy) were found in each language category, 
across both contests. This figure reports bugs before unification
during the fix-it phase, and is of course 
affected by differences among teams' skills and language choices in the two contests, 
but it provides a high-level perspective. 

Our model shows that teams that knew more unique languages
(even if they did not use those languages in their submission)
performed slightly better, about $1.2\times$ for each language
known. Additional LOC in an implementation were also associated with a
very small increase in the presence of an identified security
bug. % \michelle{did we want a graph for unique languages?}

Finally, the model shows two factors that played a role in the
outcome, but not in a statistically significant way: using a
dynamically typed language, and participating in the MOOC. We see the
effect of the former in Figure~\ref{fig:bugsfound-by-language}. For
the latter, the effect size is quite large; it is possible that the
MOOC security training played a role.

\subsection{Breaking success}
\label{ss:breaking-success}

Now we turn our attention to break-it team performance, i.e., how
effective teams were at finding defects in others' submissions.
First, we consider how and why teams performed as indicated by their
(normalized) break-it score \emph{prior to the fix-it phase}. 
We do this to measure a team's raw output, ignoring whether other
teams found the same bug (which we cannot assess with confidence
due to the lack of fix-it phase participation
per~\S\ref{sec:resilience}). 
This data set includes 108 teams that participated in the break-it
phase in Spring and Fall 2015. 
We also model which factors contributed to \textbf{security bug
  count}, or how many total security bugs a break-it team found. 
Doing this disregards a break-it team's effort at finding correctness
bugs.

%securityCount ~ contest_problem + members_count + knowC + languageCount + 
%    coding_experience + coursera + build_participant + advanced_technique

We model both break-it score and security bug count using several of the same 
potential factors as discussed previously, 
but applied to the breaking team rather than the building team. 
In particular, we include which contest they participated in, whether they were
\textbf{MOOC} participants, the number of break-it \textbf{Team members}, 
average team-member \textbf{Coding experience},
average team-member \textbf{Knowledge of C}, and unique
\textbf{Languages known} by the break-it team members. We also add two
new potential factors:

\textbf{Build participant:} Whether the breaking team also qualified during the build-it phase. 

\textbf{Advanced techniques:} Whether the breaking team reported using software analysis
or fuzzing to aid in bug finding. Teams that only used manual inspection and 
testing are categorized as false. 26 break-it teams (24\%) reported using advanced techniques.

For these two initial models, our potential factors provide eight degrees of freedom; again 
assuming power of 0.75, this yields a prospective effect size $f^2=0.136$, indicating we could 
again expect to find effects of roughly medium size by Cohen's heuristic~\cite{Cohen:1988}. 

\begin{table}[t]
\begin{center}
\small
\begin{tabular}{l r r r }
\toprule
\textbf{Factor} & \textbf{Coef.} & \textbf{SE} & \textbf{$p$-value} \\
\midrule
Fall 2015 & -2406.89 & 685.73 & <0.001* \\
\# Team members & 430.01 & 193.22 & 0.028* \\
\color{Gray} Knowledge of C & \color{Gray} -1591.02 & \color{Gray} 1006.13 & \color{Gray} 0.117~ \\
\color{Gray} Coding experience & \color{Gray} 99.24 & \color{Gray} 51.29 & \color{Gray} 0.056~ \\
\color{Gray} Build participant & \color{Gray} 1534.13 & \color{Gray} 995.87 & \color{Gray} 0.127~ \\
\bottomrule
\end{tabular}
\end{center}
\vspace{-2ex}
\caption{
Final linear regression model of teams' break-it scores, indicating how many points 
each selected factor adds to the total score. % (The full set of 
% initial factors is listed in~\S\ref{ss:breaking-success}; this model reflects minimum 
% AIC). Significant values are indicated with an asterisk.
Overall
effect size $f^2 = 0.039$.}
\label{tab:breakit-model}
\vspace{-4ex}
\end{table}

\paragraph*{Break score} The model considering break-it score is given
in Table~\ref{tab:breakit-model}. It shows that teams with more
members performed better, with an average of 430 additional points per
team member. Auditing code for errors is
an easily parallelized task, so teams with more members could divide
their effort and achieve better coverage. Recall that having more team
members did not help build-it teams (see Tables~\ref{tab:ship-model} and~\ref{tab:secBug-model});
this makes sense as development requires more coordination, especially
during the early stages.
 % (On the flip side, it is
% interesting that the same advantage was not conferred to build-it
% teams.\pxm{The reason why build-it teams did not have this advantage
%   is given in the previous sentence, so perhaps not that interesting?}) 

\begin{figure}[t!]
  \begin{subfigure}{\columnwidth}
    \centering
    \includegraphics[width=0.98\columnwidth]{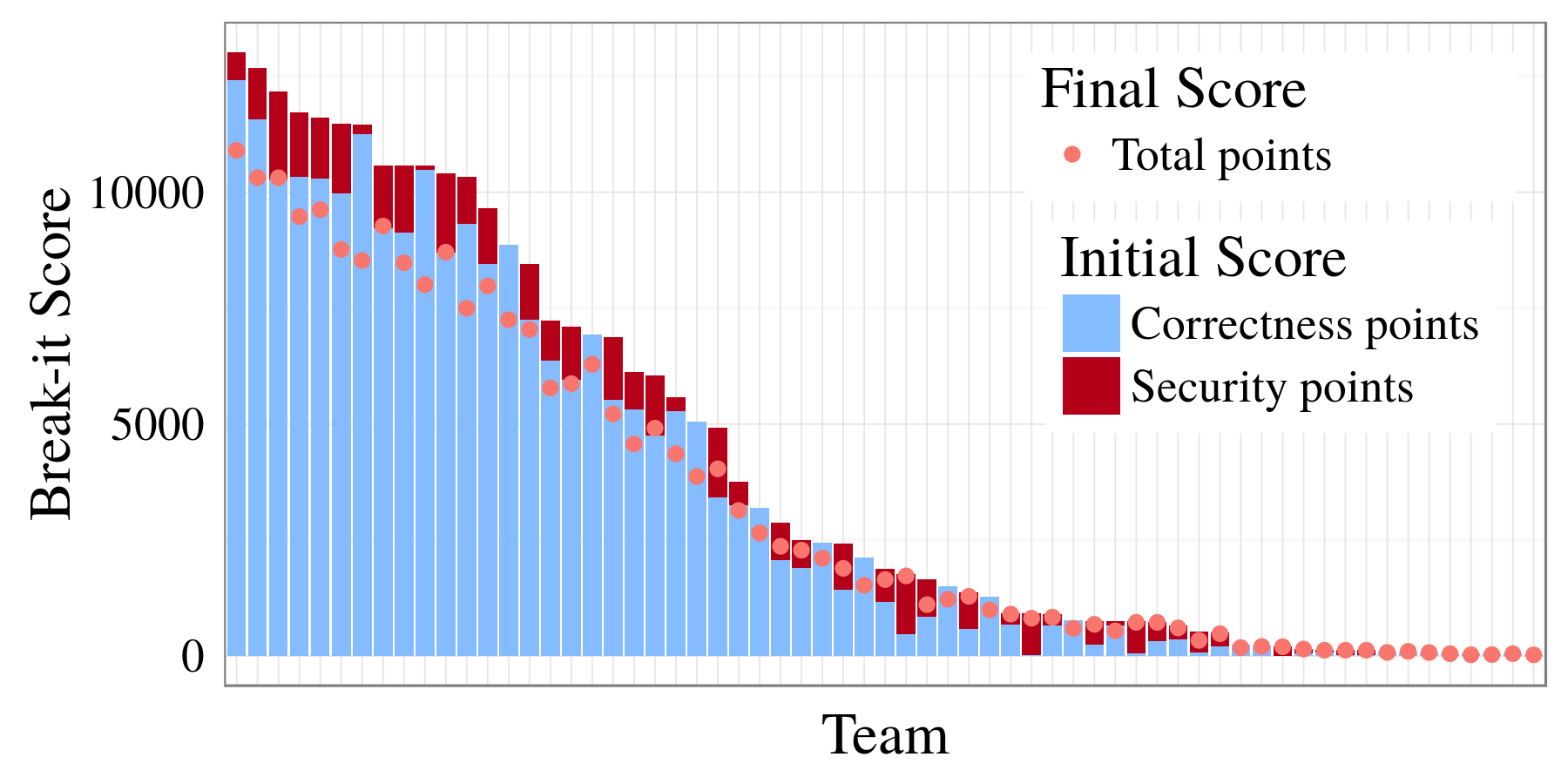}
	\caption{Spring 2015}
  \end{subfigure}
  \hfill
  \begin{subfigure}{\columnwidth}
    \centering
    \includegraphics[width=0.98\columnwidth]{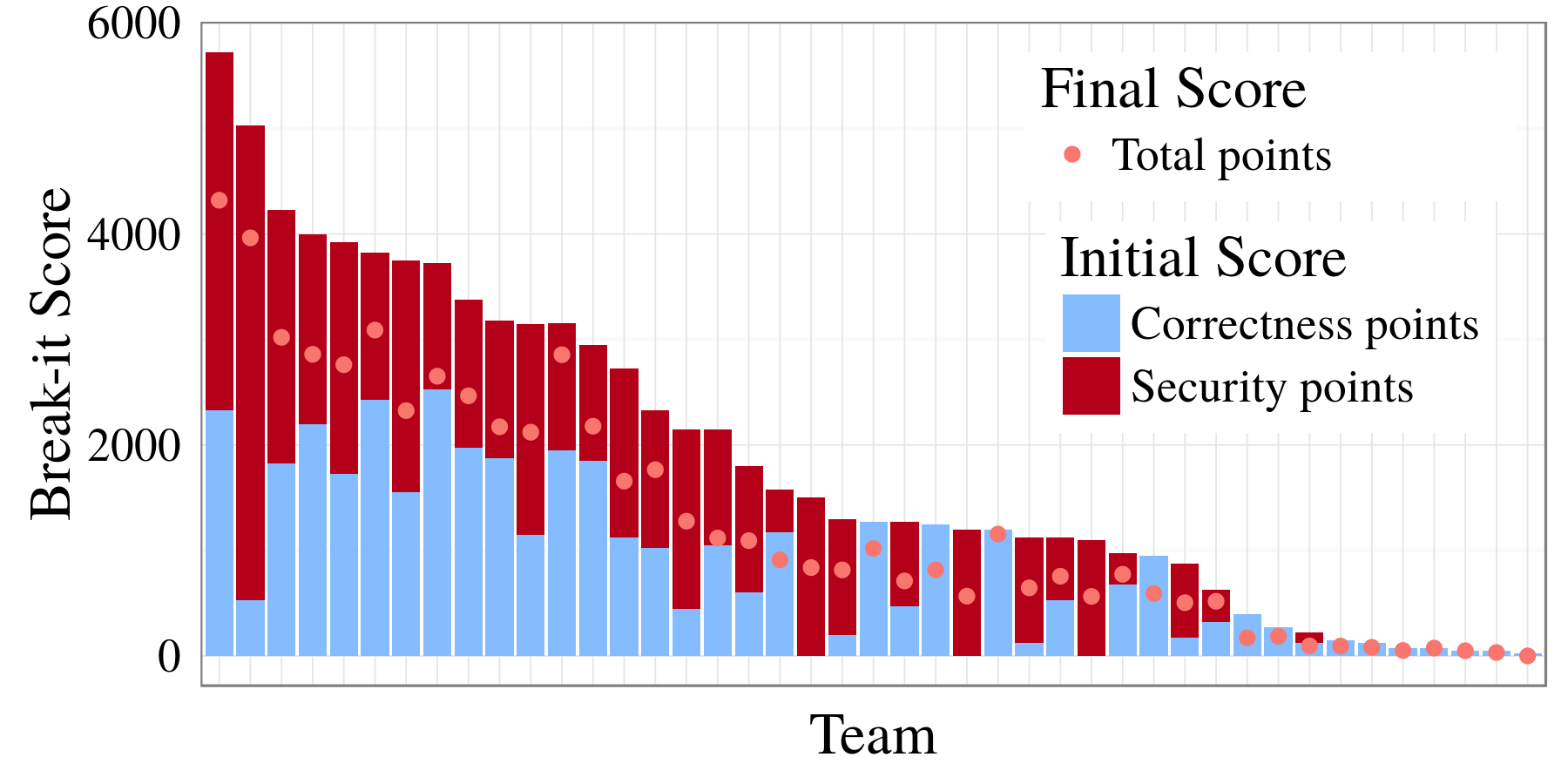}
	\caption{Fall 2015}
  \end{subfigure}
\caption{\label{fig:security-v-correctness}
Scores of break-it teams prior to the fix-it phase, broken down by points from security 
and correctness bugs. The final score of the break-it team (after fix-it phase) is noted as a dot.
Note the different ranges in the $y$-axes; in
general, the Spring 2015 contest (secure log problem) had higher scores for breaking.}
\end{figure}

The model also indicates that Spring 2015 teams performed
significantly better than Fall 2015 teams.
Figure~\ref{fig:security-v-correctness} illustrates that correctness
bugs, despite being worth fewer points than security bugs, dominate overall 
break-it scores for Spring 2015. In Fall 2015 the scores 
are more evenly distributed between correctness and security bugs.
This outcome is not surprising to us, as it was somewhat by design.
The Spring 2015 problem defines a rich command-line interface 
with many opportunities for subtle errors that break-it teams can
target. It also allowed a break-it team to submit up to 10 correctness
bugs per build-it submission. To nudge teams toward finding more
security-relevant bugs, we reduced the submission limit from 10 to 5,
and designed the Fall 2015 interface to be far simpler. 
 
Interestingly, making use of advanced analysis techniques did not
factor into the final model; i.e., such techniques did not provide a
meaningful advantage. This makes sense when we consider that such
techniques tend to find generic errors such as crashes, bounds
violations, or null pointer dereferences. Security violations for our
problems are more semantic, e.g., involving incorrect design or use of
cryptography. Many correctness bugs were non-generic too, e.g.,
involving incorrect argument processing or mishandling of inconsistent
or incorrect inputs.

Being a build participant and having more coding experience is
identified as a postive factor in the break-it score, according to the
model, but neither is statistically significant (though they are close
to the threshold). Interestingly,
knowledge of C is identified as a strongly negative factor in break-it
score (though again, not statistically significant). Looking closely
at the results, we find that \emph{lack} 
of C knowledge is extremely \emph{uncommon}, but that the handful of
teams in this category did unusually well. However, there are to few
of them for the result to be significant.

\begin{table}[t]
\begin{center}
\small
\begin{tabular}{l r r r }
\toprule
\textbf{Factor} & \textbf{Coef.} & \textbf{SE} & \textbf{$p$-value} \\
\midrule
Fall 2015 & 3.847 & 1.486 & 0.011* \\
\# Team members & 1.218 & 0.417 & 0.004* \\
Build participant & 5.430 & 2.116 & 0.012* \\
\bottomrule
\end{tabular}
\end{center}
\vspace{-2ex}
\caption{
Final linear regression modeling the count of security bugs found by each team. 
Coefficients indicate how many security bugs each factor adds to the
count. %  (The 
% full set of initial factors is listed in~\S\ref{ss:breaking-success}; this model reflects minimum 
% AIC). Significant values are indicated with an asterisk.
Overall effect size $f^2 = 0.035$.}
%}
\label{tab:securitycount-model}
\vspace{-2ex}
\end{table}

\paragraph*{Security bugs found}

\begin{figure}[t!]
\begin{center}
\includegraphics[width = 0.85\columnwidth]{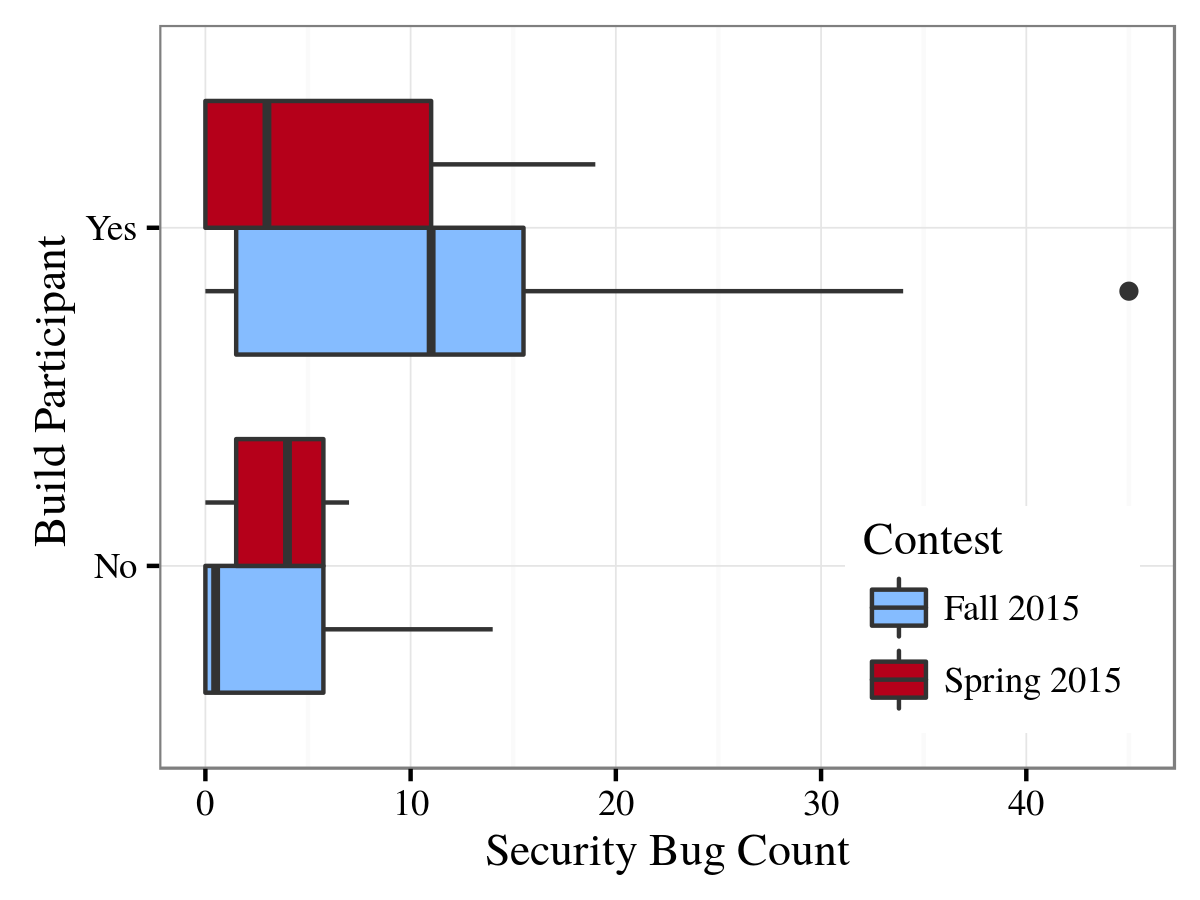}
\end{center}
\vspace{-2ex}
\caption{Count of security bugs found by each break-it team, organized by contest and 
whether the team also participated in build-it. The heavy vertical
line in the box is the median, the boxes show the first and third quartiles, and the whiskers 
extend to the most outlying data within $\pm1.5\times$ the interquartile range. 
Dots indicate further outliers.}
\vspace{-2ex}
\label{fig:secBugCount-by-builder-box}
\end{figure}

We next consider breaking success as measured by the count of security
bugs a breaking team found.  This model
(Table~\ref{tab:securitycount-model}) again shows that team size is
important, with an average of one extra security bug found for each
additional team member. 
Being a qualified builder also significantly helps one's score; this makes intuitive sense, as
one would expect to gain a great deal of insight into how a system
could fail after successfully building a similar system. 
Figure~\ref{fig:secBugCount-by-builder-box} shows the distribution of
the number of security bugs found, per contest, for break-it teams
that were and were not qualified build-it teams. 
%
%\mwh{Not sure I like the following; feel free to reword.}
Note that all but three of the 108 break-it teams made some
attempt, as defined by having made a commit, to participate during the build-it
phase---most of these (93) qualified, but 12 did not. If the reason
was that these teams were less capable programmers, that may imply
that programming ability generally has some correlation with break-it success.

%\begin{figure}[t!]
%\begin{centering}
%\includegraphics[width=\columnwidth]{figures/p13}
%\end{centering}
%\label{fig:secBugCount-by-builder-DNA}
%\caption{\dml{Figure p13 needs a caption}. Keep p12 or p13 but not both.}
%\end{figure}

On average, four more security bugs were found by a 
Fall 2015 team than a Spring 2015 team. This contrasts with the finding that Spring 2015
teams had higher overall break-it scores, but corresponds to the finding 
that more Fall 2015 submissions had security bugs found against them. 
As discussed above, this is because correctness bugs dominated in Spring 2015 
but were not as dominant in Fall 2015. Once again, the reasons may
have been the smaller budget on per-submission correctness bugs in
Fall 2015, and the greater potential attack surface in the ATM problem. 

%\michelle{say why---wrong answer in comments
%  below.} % As discussed above and
% illustrated in Figure~\ref{fig:security-v-correctness}, this is
% because correctness bugs dominated overall scores. Also as discussed
% above, we believe the higher incidence of security bugs found in
% Spring 2015 relates partially to the ease of demonstrating a security
% bug against the log problem, in contrast to writing a
% man-in-the-middle exploit for the ATM problem.

\section{Qualitative Analysis}
\label{sec:stories}

As part of the data gathered, we also have the entire program produced during 
the build-it phase as well as the programs patched during the fix-it phase. 
We can then perform a qualitative analysis of the programs which is guided by
knowing the security outcome of a given program. Did lots of break-it teams
find bugs in the program, or did they not? What are traits or characteristics 
of well-designed programs?

%\newpage
\subsection{Success Stories}
The success stories bear out some old chestnuts of wisdom in the security community:
submissions that fared well through the break-it phase made heavy use of existing 
high-level cryptographic libraries with few ``knobs'' that allow for incorrect usage\cite{bernstein2012security}.

One implementation of the ATM problem, written in Python, made use of the SSL PKI infrastructure. The
implementation used generated SSL private keys to establish a root of trust that authenticated the 
\texttt{atm} program to the \texttt{bank} program. Both the \texttt{atm} and \texttt{bank} required that the connection be signed with 
the certificate generated at runtime. Both the \texttt{bank} and the \texttt{atm} implemented their communication 
protocol as plain text then wrapped in HTTPS. This put the contestant on good footing; to 
find bugs in this system, other contestants would need to break the security of OpenSSL. 

Another implementation, also for the ATM problem, written in Java, used the NaCl library. 
This library intentionally provides a very high level API to ``box'' and ``unbox'' secret values, freeing
the user from dangerous choices. As above, to break this system, other contestants would need to 
first break the security of NaCl. 

An implementation of the log reader problem, also written in Java, achieved success using a high
level API. They used the BouncyCastle library to construct a valid encrypt-then-MAC scheme over
the entire log file. 

\subsection{Failure Stories}
The failure modes for build-it submissions are distributed along a spectrum ranging from 
``failed to provide any security at all'' to ``vulnerable to extremely subtle timing attacks.'' 
This is interesting because it is a similar dynamic observed in the software marketplace today. 

Many implementations of the log problem lacked encryption or authentication.
Exploiting these design flaws was trivial for break-it teams. Sometimes log data
was written as plain text, other times log data was serialized using the Java object serialization
protocol. 

% The nature of the log problem, and the break it infrastructure, limited some options available
% to breakers. 
% One break it team contacted us to say that, if they could interact with a particular
% implementation 50 times, they could reveal confidential information. 
% Our framework did not support
% this type of interactive attack, so it was not counted against the
% build it team.
One break-it team discovered a privacy flaw which they could exploit with at most fifty probes. 
The target submission truncated the ``authentication token,'' so that it was vulnerable to a brute force attack. 

The ATM problem allows for interactive attacks (not possible for the log), and the attacks became cleverer 
as implementations used cryptographic constructions incorrectly. One implementation used cryptography,
but implemented RC4 from scratch and did not add any randomness to the key or the cipher stream. An 
attacker observed that the ciphertext of messages was distinguishable and largely unchanged from 
transaction to transaction, and was able to flip bits in a message to change
the withdrawn amount. 

Another implementation used encryption with authentication, but did
not use randomness; as such error messages were always distinguishable
success messages. An attack was constructed against 
this implementation where the attack leaked the bank balance by observing different withdrawal
attempts, distinguishing the successful from failed transactions, and performing a binary search to identify
the bank balance given a series of withdraw attempts. 

Some failures were common across ATM problem implementations. Many implementations kept the key fixed
across the lifetime of the \texttt{bank} and \texttt{atm} programs and did not use a nonce in the messages. 
This allowed attackers to replay messages freely between the
\texttt{bank} and the \texttt{atm}, violating integrity via
unauthorized withdrawals. Several implementations used encryption, but without authentication. 
These implementations used a library such as OpenSSL, the Java cryptographic framework, or the Python
pycrypto library to have access to a symmetric cipher such as AES, but
either did not use these libraries at a level where authentication was provided in addition to encryption,
or they did not enable authentication. 

Some failures were common across log implementations as well: if an implementation used 
encryption, it might not use authentication. If it used authentication, it would authenticate
records stored in the file individually and not globally. The implementations would also
relate the ordering of entries in the file to the ordering of events in time, allowing for
an integrity attack that changes history by re-ordering entries in the file. 

As a corpus for research, this data set is of interest for future mining. What common design
patterns were used and how did they impact the outcome? Are there any metrics we can extract
from the code itself that can predict break-it scores? We defer this analysis
to future work.

\section{Related work}
\label{sec:related}

\bibifi bears similarity to existing programming and security
contests but is unique in its focus on building secure
systems. \bibifi also is related to studies of code and secure
development, but differs in its open-ended contest format.

\paragraph*{Contests} Cybersecurity contests typically
focus on vulnerability discovery and exploitation, and sometimes involve a system
administration component for defense.  
% This approach of ``gamification'' of security is attractive; Gary
% McGraw calls this the NASCAR effect~\cite{mcgraw12active}. No
% existing public contest at scale focuses on secure development.
%
One popular style of contest is dubbed \emph{capture the flag} (CTF)
and is exemplified by a contest held at DEFCON~\cite{defcon}. Here,
teams run an identical system that has buggy components. The goal is
to find and exploit the bugs in other competitors' systems while
mitigating the bugs in your own. Compromising a system enables a team
to acquire the system's key and thus ``capture the flag.'' In addition to
DEFCON CTF, there are other CTFs such as iCTF~\cite{Childers2010hacking,Doupe2011live} 
and PicoCTF~\cite{picoctf}. The use of this style of contest in an educational setting
has been explored in prior work~\cite{conti2011competition,eagle2013competition,hoffman05exercise}.
The Collegiate Cyber Defense Challenge~\cite{nationalccdc,conklin2006cyber,Conklin2005ccdc} and
the Maryland Cyber Challenge \& Competition~\cite{mdc3} have
contestants defend a system, so their
responsibilities end at the identification and mitigation of
vulnerabilities. These contests focus on bugs in systems as a key
factor of play, but neglect software development.

Programming contests challenge students to build clever, efficient
software, usually with constraints and while under (extreme) time
pressure. The ACM programming contest~\cite{acmprogramming} asks teams
to write several programs in C/C++ or Java during a 5-hour time
period. Google Code Jam~\cite{codejam} sets tasks that must be solved
in minutes, which are then graded according to development speed
(and implicitly, correctness). Topcoder~\cite{topcoder} runs
several contests; the Algorithm competitions are small projects that
take a few hours to a week, whereas Design and Development
competitions are for larger projects that must meet a broader
specification.  Code is judged for correctness (by passing tests),
performance, and sometimes subjectively in terms of code quality or
practicality of design.  All of these resemble the build-it phase of
\bibifi but typically consider smaller tasks; they do not consider the
security of the produced code.

\paragraph*{Studies of secure software development}

There have been a few studies of different methods and techniques for
ensuring security. Work by Finifter and
Wagner~\cite{finifter11exploring} and
Prechelt~\cite{prechelt2011plat_forms} relates to both our build-it
and break-it phases: they asked different teams to develop
the same web application using different frameworks, and then
subjected each implementation to automated (black box) testing and
manual review. They found that both forms of review were effective in
different ways, and that framework support for mitigating certain
vulnerabilities improved overall security. Other studies focused on the
effectiveness of vulnerability discovery techniques, e.g., as might be
used during our break-it phase. Edmundson et al.~\cite{codereview}
considered manual code review; Scandariato et
al.~\cite{scandariato2013static} compared different vulnerability
detection tools; other studies looked at software properties that might
co-occur with security problems
\cite{DBLP:conf/issre/WaldenSS14,yang2016improving,
  harrison2010empirical}. \bibifi differs from all of these in its
open-ended, contest format: Participants can employ any technique they
like, and with a large enough population and/or measurable impact, the
effectiveness of a given technique will be evident in final outcomes.

\section{Conclusions}\label{sec:conclusions}

This paper has presented Build-it, Break-it, Fix-it (\bibifi), a new security
contest that brings together features from typical security contests,
which focus on vulnerability detection and mitigation but not secure
development, and programming contests, which focus on development but
not security.  During the first phase of the contest, teams construct
software they intend to be correct, efficient, and secure. During the
second phase, break-it teams report security vulnerabilities and other
defects in submitted software. In the final, fix-it, phase, builders
fix reported bugs and thereby identify redundant defect reports. Final
scores, following an incentives-conscious scoring system, reward the
best builders and breakers. 

During 2015, we ran three contests
involving a total of 116 teams and two different programming
problems. Quantitative analysis from these contests found that the
best performing build-it submissions used C/C++, but submissions coded
in a statically-typed language were less likely to have a security
flaw; build-it teams with diverse programming-language knowledge also
produced more secure code. Shorter programs correlated with better
scores. Break-it teams that were also successful build-it 
teams were significantly better at finding security bugs.

There are many interesting areas of future work that BIBIFI opens up.
The BIBIFI design lends itself well to conducting more focused studies;
our competitions allow participants to use any languages and
tools they desire, but one could narrow their options for closer
evaluation.
Although we have reduced manual judging considerably, an interesting
technical problem that often arises is determining whether two
bugs are morally equivalent; an automated method for determining this
could be broadly applicable.
Finally, one limitation of our study is that we do not evaluate whether
break-it teams find all of the bugs there are to find; one improvement
would be to apply a set of fuzzers and static analyzers, or to recruit
professional teams to effectively participate in the break-it phase as
a sort of baseline against which to compare the break-it teams'
performance.

We plan to freely release \bibifi to support future research. We
believe it can act as an incubator for ideas to improve secure
development. More information, data, and opportunities to participate
are available at \url{https://builditbreakit.org}.

\section*{Acknowledgments} We thank Jandelyn Plane and Atif Me\-mon
who contributed to the initial development of BIBIFI and its
preliminary data analysis. Many people in the security community, too
numerous to list, contributed ideas and thoughts about BIBIFI during
its development---thank you! Bobby Bhattacharjee and the anonymous
reviewers provided helpful comments on drafts of this paper. This
project was supported gifts from Accenture, AT\&T, Galois, Leidos,
Patriot Technologies, NCC Group, Trail of Bits, Synposis, ASTech
Consulting, Cigital, SuprTek, Cyberpoint, and Lockheed Martin; by a
grant from the NSF under award EDU-1319147; by DARPA under contract
FA8750-15-2-0104; and by the U.S.  Department of Commerce, National
Institute for Standards and Technology, under Cooperative Agreement
70NANB15H330.

\begin{small}
\bibliographystyle{acm}
\bibliography{confs_long,proposal,atif,mwh,dml}
\end{small}

\end{document}